\documentclass[a4paper]{article}
\usepackage{graphicx}
\usepackage{amsmath,amssymb}

\newcommand\beq{\begin{equation}}
\newcommand\eeq{\end{equation}}
\newcommand\bea{\begin{eqnarray}}
\newcommand\eea{\end{eqnarray}}
\title{\bf Prepotential formulation of SU(3) lattice gauge theory}
\author{Ramesh Anishetty\footnote{ramesha@imsc.res.in}\\{\it The Institute of
Mathematical Sciences,}\\ \vspace{4mm}{\it CIT-Campus, Taramani,
Chennai 600113, India}\\
Manu Mathur\footnote{manu@bose.res.in} and Indrakshi
Raychowdhury\footnote{indrakshi@bose.res.in}\\{\it S. N. Bose National Centre
for Basic Sciences,}\\{\it JD Block, Sector III, Salt Lake City, Kolkata-
700098, India.   }} 
\begin{document}
\maketitle 




\begin{abstract}
\noindent The SU(3) lattice gauge theory is reformulated in terms of SU(3) prepotential 
harmonic oscillators. This reformulation has enlarged $SU(3)\otimes U(1) \otimes U(1)$ 
gauge invariance under which the prepotential operators transform like matter fields. 
The Hilbert space of SU(3) 
lattice gauge theory is shown to be equivalent to the Hilbert space of the prepotential 
formulation satisfying certain color invariant Sp(2,R) constraints. The SU(3) irreducible 
prepotential operators which solve these Sp(2,R) constraints are used to construct SU(3) 
gauge invariant Hilbert spaces at every lattice site in terms of SU(3) gauge invariant 
vertex operators. The electric fields and the link operators are reconstructed 
in terms of these SU(3) irreducible prepotential 
operators. We show that all the SU(3) Mandelstam constraints 
become local and take very simple form within this approach. 
We also discuss the construction of all possible linearly independent 
SU(3) loop states which solve the Mandelstam constraints. The techniques 
can be easily generalized to SU(N). 

\end{abstract}
\section{Introduction}

The reformulation of gauge theories in terms of gauge invariant Wilson loops and 
strings carrying fluxes of the corresponding gauge group is an old problem in  
quantum field theory \cite{man,wil,pol,migdal,kogut,gambini}. The motivation to go from colored gluons and quarks to 
colorless loops and string degrees of freedom comes from the expectation that the 
latter framework is better suited to analyze and understand long distance non-perturbative 
issues like color confinement in QCD.  Infact, the lattice formulation of gauge theories was 
a step in this direction where one directly works with link operators (instead of gauge connections) 
which create and destroy abelian or non-abelian loop fluxes on lattice links.  
However, the two major obstacles in this loop, string approach to QCD are the non-locality 
and proliferation of loops and string states \cite{quote}. The non-locality is obvious as the loops and strings can be of 
any shapes and sizes. The problem of proliferation exists because the set of all Wilson loop states 
forms a highly over complete basis. This is because not all loop  states are mutually independent (see section 
3.3 and 4.6). Their relationships are expressed by the Mandelstam constraints. The Mandelstam constraints, in turn, 
are difficult to solve because of their non-locality (section 3.3 and 4.6). 
{\it Therefore, it is important to explore new descriptions 
of QCD where the loop, string states and their dynamics as well as the associated Mandelstam constraints 
can be analyzed locally.} As shown in 
\cite{mm,mm1}, the prepotential approach to lattice gauge theories provides such a platform. 
More precisely, this approach allows us to analyze and solve the Mandelstam constraints 
locally at each lattice site without all the irrelevant non-local details associated with the 
loop states (section 3.3 and 4.6). Towards this goal, a complete analysis was  carried out for SU(2) 
lattice gauge theory and all mutually independent loop states were constructed 
in terms of prepotential operators in \cite{mm,mm1}. 
The purpose and motivation of this work is to analyze lattice QCD or SU(3) lattice gauge theory within 
the prepotential framework. As we will see, there are many new issues which come up due to very different 
flux properties of SU(3) and SU(2) lattice gauge theories. 

\noindent The prepotential operators are harmonic oscillators belonging to the fundamental representations 
of the gauge group. Further, unlike link operators which create and destroy fluxes on the links, the 
prepotential operators are associated with the sites and create or destroy smallest units of group fluxes 
at the corresponding lattice sites. In the case of SU(2) lattice gauge theory \cite{mm}, the    
prepotential approach enabled us to cast all the SU(2) Mandelstam constraints in their 
local form. Further, all possible  mutually orthonormal loop states were explicitly  
constructed in terms of the prepotential operators. The dynamics of these orthonormal SU(2) 
loop states was shown to be governed by 3-nj Wigner coefficients. Infact, similar results 
have been obtained in the context of duality transformations in SU(2) lattice gauge theories 
in \cite{sharat1,sharat2,robson,kolawa,pietri}. More precisely, the SU(2) gauge invariant basis 
labeled by (dual) angular momentum quantum numbers, describing two dimensional triangulated 
surfaces, in \cite{sharat2} is exactly same as the SU(2) loop basis in \cite{mm} labeled by 
``linking quantum numbers" which describe one dimensional loops. In \cite{gambini,brugmann,loll,watson} 
different computational schemes to identify independent SU(2) loops were proposed. 
In \cite{gambini,brugmann} loop Hamiltonians are computed in the above schemes retaining small loops 
carrying small fluxes\footnote{Note that the prepotential formulation resolve the issues 
of over completeness of SU(2) loop states and their dynamics exactly without any assumptions.}. 
In the  context of loop quantum gravity, SU(2) spin networks carrying SU(2) fluxes which describe 
geometry of space time have been extensively studied \cite{rovelli,smolin}. The SU(2) Schwinger boson 
or equivalently  prepotential techniques studied in \cite{mm,mm1} can  also be naturally applied to study 
the spin networks in loop quantum gravity as the fluxes in the spin networks are created by Schwinger bosons. 
This approach leads to many technical simplifications in the construction of spin networks 
and has been discussed extensively in \cite{dass}.  
On the other hand, in the context of QCD with SU(3) gauge group hardly any work has been done 
in these directions. 
In particular, it is important to construct and analyze all independent 
SU(3) loop states (``SU(3) spin networks") and study their dynamics. 
In the context of QCD, this analysis will be useful 
to analyze the spectrum of QCD Hamiltonian in terms of loops near the continuum limit where large loops 
carrying large fluxes are expected to dominate. The exact minimal loop basis containing arbitrarily large loops 
with all possible fluxes will allow us to analyze the spectrum without any spurious loop degrees of freedom.  

In this work we show that 
the SU(3) lattice gauge theory can also be completely 
described in terms of SU(3) irreducible prepotentials with 
$SU(3) \otimes U(1) \otimes U(1)$ gauge invariance. Under $SU(3) \otimes U(1) \otimes U(1)$ 
gauge transformations the prepotentials transform like charged matter fields. All the 
non-local SU(3) Mandelstam constraints in term of the link operators are cast into 
their local forms with the help of SU(3) gauge invariant prepotential vertex operators 
which are defined at lattice sites (section 3.3 and 4.6). We briefly discuss how to get 
all the solutions of SU(3) Mandelstam constraints in the form of all possible independent 
SU(3) loop states.

The paper is organized as follows. In Section 2, we briefly discuss the Hamiltonian 
formulation of SU(N) lattice gauge theory. This section sets up the notations and  
makes the paper self contained. The section 3 starts with a brief summary of the 
SU(2) prepotential approach to lattice gauge theory \cite{mm,mm1}.  
This overview illustrates all the essential ideas involved in simplifying the  
Mandelstam constraints and getting all their solutions in the simpler SU(2) case 
before dealing with their more involved SU(3) analogues. 
In addition, this section also helps us to highlight some completely new issues and difficulties one confronts on 
going from SU(2) to SU(3) lattice gauge theory. Section 4 discusses  SU(3) lattice gauge theory in terms of 
prepotential operators. In sections 4.1 and 4.2  
we study and classify the SU(3) prepotential Hilbert space ${\cal H}_p$ according to  
SU(3) invariant Sp(2,R) quantum numbers \cite{nm}. In section 4.3, we show 
that the Hilbert space of SU(3) gauge theory ${\cal H}_g$ is a tiny subspace of ${\cal H}_p$ which 
satisfies certain Sp(2,R) constraints. Section 4.4 deals with SU(3) irreducible prepotential operators 
\cite{ima} which are solutions of the above Sp(2,R) constraints and therefore directly create the 
gauge theory Hilbert space ${\cal H}_g$. 
The explicit construction of $SU(3)$ link operators and electric fields in terms of 
the SU(3) irreducible prepotentials is given in section 4.5. 
In section 4.6, with the help of SU(3) irreducible prepotential operators, we construct all possible 
SU(3) gauge invariant vertices at a given lattice site which in turn   cast all SU(3) Mandelstam constraints in 
their local forms.  Having made them local, section 4.6.1 discusses how to solve these infinite sets of 
constraints at every lattice site 
exactly. We then briefly discuss the prepotential formulation of SU(N) lattice gauge theory.
We end the  paper with a brief summary and discussion on related issues. 

\section{SU(N) Hamiltonian formulation}
The Hamiltonian of $SU(N)$ lattice gauge theory is:
\bea
\hskip -1.2cm H = \sum_{n,i}\sum_{\mathrm a=1}^{N^2-1}E^{\mathrm a}(n,i) E^{\mathrm a}(n,i) 
+ K \sum_{\mbox{plaquette}} Tr \left(U_{\mbox{plaquette}} + U^{\dagger}_{\mbox{plaquette}}\right)
\label{ham}
\eea
with, 
\bea
U_{\mbox{plaquette}}=U(n,i)U(n+i,j)U^{\dagger}(n+j,i)U^{\dagger}(n,j), \nonumber 
\eea 
where $K$ is the coupling constant, 
$\mathrm a (=1,2, \cdots ,(N^2-1))$ is the color index. 
In (\ref{ham}) the kinematical operators $E$ and $U$ can be understood as follows. 
Each link (n,i) is associated with a SU(N) symmetric top, whose configuration (i.e the rotation matrix
from space fixed to body fixed frame) is given by the operator valued $(N \times N)$ 
SU(N) matrix U(n,i). 
Let $E_L^\mathrm a(n,i), E_R^\mathrm a(n+i,i)$ denote the conjugate left and right electric fields 
with the quantization rules \cite{kogut}:
\bea
\left[E_L^{\mathrm a}(n,i),U^{\alpha}{}_{\beta}(n,i)\right]  =  - \left(T^{\mathrm a}U(n,i)\right)^{\alpha}{}_{\beta},~~     
~~~\left[E_R^{\mathrm a}(n+i,i),U^{\alpha}_{\beta}(n,i)\right]  =  \left(U(n,i) T^\mathrm a \right)^{\alpha}{}_{\beta}.    
\label{ccr} 
\eea 
In (\ref{ccr}),  $T^{\mathrm a}$ are the  generators in the fundamental representation of 
$SU(N)$ and satisfy: $[T^{\mathrm a},T^{\mathrm b}]=if^{\mathrm{abc}}T_{\mathrm c}$  where 
$f^{\mathrm {abc}}$ are the SU(N) structure constants. The quantization rules (\ref{ccr}) 
clearly show that $E_L(n,i)$ and $E_R(n+i,i)$ are the generators of left and the right 
gauge transformations in  (\ref{eugt}).  
Infact, the right generators $E^\mathrm a_R(n+i,i)$ are the 
parallel transport of the left generator $E^\mathrm a_L(n,i)$  on the link $(n,i)$:  
\bea
\label{Ee}
\hskip 2cm E_R(n+i,i)= - U^{\dagger}(n,i) E_L(n,i) U(n,i). 
\eea
In (\ref{Ee}), $E_R(n+i,i) \equiv \sum_{\mathrm a} E^\mathrm a_R(n+i,i)T^\mathrm a$ and $E_L(n,i) \equiv \sum_\mathrm a E^\mathrm a_L(n,i)T^\mathrm a$.
The left and the right electric fields on every link, being the SU(N) rotation generators, 
satisfy: 
\bea
[E_L^{\mathrm a}(n,i),E_L^{\mathrm b}(n,i)]=if_{\mathrm{abc}}E_L^{\mathrm c}(n,i), ~~ 
[E_R^{\mathrm a}(n,i),E_R^{\mathrm b}(n,i)]=if_{\mathrm{abc}}E_R^{\mathrm c}(n,i).
\label{comee}
\eea
Further, using (\ref{Ee}), it is easy to show that $E_L^\mathrm a$ and $E_R^\mathrm a$ commute amongst themselves: 
\bea 
\left[E_L^\mathrm a(n,i),E_R^\mathrm b(m,j)\right]=0.
\label{cz}
\eea 
and therefore mutually independent. 
By construction on each link they always satisfy the constraints:
\bea
\label{E2=e2}
\sum_{{\mathrm a}=1}^{N^2-1}E^{\mathrm a}(n,i)E^{\mathrm a}(n,i) \equiv 
\sum_{{\mathrm a}=1}^{N^2-1}E_L^{\mathrm a}(n,i)E_L^{\mathrm a}(n,i)
=\sum_{{\mathrm a}=1}^{N^2-1}E_R^{\mathrm a}(n+i,i)E_R^{\mathrm a}(n+i,i).
\eea
The Hamiltonian in (\ref{ham}) involves the squares of either left or the 
right electric fields.  Under gauge transformation the left electric field and 
the link operator transforms as:
\bea
\label{eugt}
U(n,i)\rightarrow \Lambda(n)U(n,i)\Lambda^{\dagger}(n+i),  \hskip 4cm  
\nonumber \\ 
E_L(n,i)\rightarrow \Lambda(n)E_L(n,i)\Lambda^{\dagger}(n),~~
E_R(n+i,i)\rightarrow \Lambda(n+i)E_R(n+i,i)\Lambda^{\dagger}(n+i).
\eea
The Hamiltonian (\ref{ham}) and the basic commutation relations (\ref{ccr}) are  invariant under 
the SU(N) gauge transformations (\ref{eugt}). From (\ref{eugt}),  the $SU(N)$ Gauss law 
constraint at every lattice site $n$ is
\bea
\label{gl}
\hskip 2cm G(n) = \sum_{i=1}^{d}\Big( E_L^{\mathrm a}(n,i) + E_R^{\mathrm a}(n,i)\Big)=0, \forall n. 
\eea
It is convenient to define the left and right strong coupling vacuum state $|0\rangle_L$ and $|0\rangle_R$ 
on every link which are annihilated by their corresponding electric fields: 
\bea 
E_L^\mathrm a(n,i)|0, (n,i)\rangle_L  = 0, ~~~~  E_R^\mathrm a(n+i,i)|0, (n+i,i) \rangle_R = 0, ~~ \forall~{\textrm{links}} ~(n,i).
\label{scv}
\eea   
We will  denote the vacuum state on a link by $|0\rangle \equiv |0, (n,i)\rangle_L \otimes |0, (n,i)\rangle_R$, 
suppressing all the link as well as L, R indices. The quantization rules (\ref{ccr}) show that the 
link operators $U^{\alpha}{}_\beta(n,i)$ acting on the strong coupling vacuum (\ref{scv}) create  SU(N) 
fluxes on the links. As an example, using (\ref{ccr}): 
\bea 
E_L^2(n,i) \Big(U^{\alpha}{}_\beta |0\rangle\Big)    = 
E_R^2(n+i,i) \Big(U^{\alpha}{}_\beta| 0 \rangle\Big)   = \frac{1}{2N}\left(N^2-1\right) 
\Big(U^{\alpha}{}_\beta| 0 \rangle\Big). 
\label{feq}
\eea 
The higher SU(3) irreducible flux eigenstates of $E_L^2$ and $E_R^2$ on a link can be obtained by considering 
the states $U^{\alpha_1}{}_{\beta_1} U^{\alpha_2}{}_{\beta_2}  \cdots U^{\alpha_1}{}_{\beta_1}  
|0\rangle$ and symmetrizing the $\alpha$ and therefore also $\beta$ indices  according to certain 
SU(N) Young tableau.  We will discuss this issue again in section 3 and section 4 in the specific 
context  of SU(2) and SU(3) groups.  

\section{Prepotentials in SU(2) lattice gauge theory}

\begin{figure}[h]
\begin{center}
\includegraphics[width=0.7\textwidth,height=0.15\textwidth]
{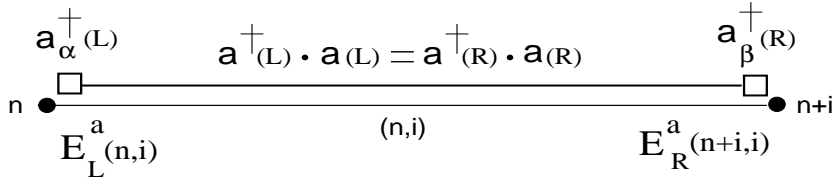}
\end{center}
\caption{The left and right  electric fields and the corresponding prepotentials in SU(2) lattice gauge theory.
We have denoted $a^{\dagger}(n,i,L)$ and $a^{\dagger}(n+i,i,R)$ by $a^{\dagger}(L)$ and  $a^{\dagger}(R)$ 
respectively.  The  unoriented abelian flux line  connecting them  represents the U(1) Gauss law (\ref{noc}) constraint.}  
\label{su2pp}
\end{figure}

In this section we define SU(2) prepotential operators.
Using the Schwinger bosons construction of the angular momentum algebra (\ref{comee}), the 
left and the right electric fields on a link $(n,i)$ can be written as: 
\bea
\label{sb} 
\mbox{Left electric fields:} ~~ \quad \quad \quad E_L^{\mathrm a}(n,i) &\equiv&  
a^{\dagger}(n,i;L)\frac{\sigma^{\mathrm a}}{2}a(n,i;L),  \\ 
\mbox{Right electric fields:} \quad ~ E_R^{\mathrm a}(n+i,i) &\equiv & 
a^{\dagger}(n+i,i;R)\frac{\sigma^{\mathrm a}}{2}a(n+i,i;R). \nonumber 
\eea
In (\ref{sb}), $a_{\alpha}(n,i;l)$ and  $a_{\alpha}^{\dagger}(n,i;l)$ are the  doublets of harmonic 
oscillator creation and annihilation operators with $l= L,R, \alpha =1,2$. We have used Schwinger 
boson construction \cite{schwinger} of angular momentum algebra in (\ref{sb}). 
Like $E_L^\mathrm a(n,i)$  and $E_R^\mathrm a(n+i,i)$, the locations of $a(n,i,L),a^{\dagger}(n,i,L)$ 
and $a(n+i,i,R),a^{\dagger}(n+i,i,R)$  are on the left and the right of the link 
$(n,i)$. For notational convenience we suppress the link indices and 
denote $a^{\dagger}(n,i,L)$ and $a^{\dagger}(n+i,i,R)$ by $a^{\dagger}(L)$ and $a^{\dagger}(R)$ 
respectively.  This is clearly illustrated in Figure \ref{su2pp}.  
The link indices will be explicitly shown whenever we work with more than one link. 
Note that the relations (\ref{sb}) imply that the strong coupling vacuum (\ref{scv}) is 
the harmonic oscillator vacuum. Substituting the electric fields (\ref{sb}) in terms of 
Schwinger bosons in the electric field constraints (\ref{E2=e2}), we get 
$ a^{\dagger}(n,i;L) \cdot a(n,i;L) =  a^{\dagger}(n+i,i;R) \cdot a(n+i,i;R)$. We will 
come back to this issue again in section 3.1.  

\noindent Under SU(2) gauge transformation with the generator G(n) in (\ref{gl}), the prepotential harmonic 
oscillator transform as SU(2) doublets\footnote{Here we specify the notations in (\ref{su2gt}):   
$a^{\dagger}_{\alpha}(L) \equiv a^{\dagger}_\alpha(n,i;L), a^{\dagger}_{\alpha}(R) \equiv a^{\dagger}_\alpha(n+i,i;R)$ 
are located at the left and right side of the link $(n,i)$ and $\Lambda_L \equiv \Lambda(n), ~ \Lambda_R \equiv 
\Lambda(n+i)$ as shown in Figure \ref{su2pp} explicitly.}: \bea 
a^{\dagger}_{\alpha}(L) \rightarrow a^{\dagger}_{\beta}(L)~ \big(\Lambda_L^{\dagger}\big)^{\beta}{}_{\alpha}, \quad \quad  
a^{\dagger}_{\alpha}(R) \rightarrow a^{\dagger}_{\beta}(R)~ \big(\Lambda_R^{\dagger}\big)^{\beta}{}_{\alpha} \nonumber \\
a^{\alpha}(L) \rightarrow \big(\Lambda_L\big)^{\alpha}{}_{\beta}~ a^{\beta}(L), \quad \quad  
a^{\alpha}(R) \rightarrow \big(\Lambda_R\big)^{\alpha}{}_{\beta}~  a^{\beta}(R).  
\label{su2gt} 
\eea 
One can also define $\tilde a^{\dagger\alpha}=\epsilon^{\alpha\beta}a^{\dagger}_{\beta}$ and $\tilde 
a_\alpha=\epsilon_{\alpha\beta}a^{\beta}$ which under $SU(2)$ transformation transform as $a^\alpha$ and 
$a^{\dagger}_\alpha$ respectively. 
In terms of link operators the basic SU(2) flux states on links can be constructed using the link 
operators: 
\bea 
|j(n,i), m_L(n,i), m_R(n,i)\rangle
=  \Big(U^{\alpha_1}{}_{\beta_1} U^{\alpha_2}{}_{\beta_2} \cdots U^{\alpha_{2j}}{}_{\beta_{2j}} + 
\cdots {\textrm{(2j)! permutations}} \Big) |0\rangle.  
\label{ubasis} 
\eea
In (\ref{ubasis}), $j_L(n,i) = j_R(n+i,i) \equiv j(n,i)$ because of (\ref{E2=e2}), 
$m_L = \sum_{i=1}^{2j} \alpha_i$ and $m_R = \sum_{i=1}^{2j} \beta_i$ with 
$\alpha_i,\beta_i = \pm \frac{1}{2}$. 
The $(2j)!$ terms in (\ref{ubasis}) are required to implement 
the symmetries of SU(2) Young tableau  
in the left ($\alpha_1 \alpha_2 \cdots \alpha_{2j}$) as well as the right ($\beta_1 \beta_2 \cdots 
\beta_{2j}$) indices. 
The gauge theory Hilbert space 
${\cal H}_g$ is spanned by direct product of states of type (\ref{ubasis}) on 
all the lattice links.  Note that as the flux value $j \rightarrow \infty$ on various links\footnote{These 
large j configurations are expected to dominate in the continuum ($g \rightarrow 0$) limit.}, 
the construction of the gauge theory Hilbert space ${\cal H}_g$ through (\ref{ubasis}) becomes 
more and more tedious.  
The basic link states in (\ref{ubasis}) can be now be disentangled into it's left and 
right part as: 
\bea 
|j(n,i), m_L(n,i), m_R(n,i)\rangle  =  |j(n,i),m_L(n,i)\rangle_L \otimes |j(n,i),m_R(n,i)\rangle_R, 
\label{lrp} 
\eea 
 where,   
\bea 
|j(n,i),m_L(n,i)\rangle_L  & = &   
a^{\dagger}_{\alpha_1}(L) a^{\dagger}_{\alpha_2}(L)\cdots a^{\dagger}_{\alpha_{n}}(L) |0\rangle_L ~\equiv ~ 
\hat{{\cal{L}}}_{\alpha_1\alpha_2 \cdots \alpha_n}  |0\rangle_L,  \nonumber \\ 
|j(n,i),m_R(n,i)\rangle_R & =  &  
a^{\dagger}_{\beta_1}(R) a^{\dagger}_{\beta_2}(R) \cdots a^{\dagger}_{\beta_{n}}(R) |0\rangle_R ~\equiv~ 
\hat{{\cal{R}}}_{\beta_1\beta_2 \cdots \beta_n}  
|0\rangle_R.  
\label{pls} 
\eea 
\noindent In (\ref{pls}), $n =2j, ~m_L = \sum_{i=1}^{2j} \alpha_i$ and $m_R = \sum_{i=1}^{2j} \beta_i$ with 
$\alpha_i,\beta_i =\pm \frac{1}{2}$. The operators $\hat{{\cal{L}}}$ and $\hat{{\cal{R}}}$ are the 
$SU(2) \otimes U(1)$ flux creation operators at the left and right end of every link. Note that these 
operators are SU(2) irreducible 
as they are symmetric in all the SU(2) spin half indices  and are defined for later 
convenience (see section 4.2). From (\ref{ubasis}) 
and (\ref{pls}) we conclude that the Hilbert space ${\cal H}_p$ created using the prepotential 
operators on all lattice links is also the SU(2) gauge theory Hilbert space: 
\bea 
{\cal H}_g \equiv {\cal H}_p.
\label{hgp}
\eea 
However, the construction of ${\cal H}_g$ using the prepotentials (\ref{pls}) is much simpler than the equivalent  
equivalent construction (\ref{ubasis}) using  the link operators. This simplicity occurs  because  unlike 
the link operators $U_{\alpha\beta}(n,i)$ which are associated with links, the prepotential operators are  
attached to the sites (i.e, left or right ends of every link).   
Further,  all the SU(2) prepotential creation operators commute amongst themselves and we do not need 
$(2j)!$ terms (as in (\ref{ubasis})) to get the symmetries of SU(2) Young tableau.  
In words, the symmetries of SU(2) Young tableau are inbuilt in SU(2) prepotential operators. 
We will come back to this symmetry issue (end of section 4.3) and the identification of ${\cal H}_g$ 
with ${\cal H}_p$ (\ref{hgp}) (see eqns. (\ref{gssp}) and (\ref{ginp})) again when we discuss SU(3) 
lattice gauge theory in terms of prepotential operators. 

\subsection{\bf $U(1)$ gauge invariance} 

The defining equations for the prepotential operators are 
invariant under $U(1) \otimes U(1)$ gauge transformations on every link: 
\bea 
a^{\dagger}_{\alpha}(L) \rightarrow e^{i\theta(L)} a^{\dagger}_{\alpha} (L),   
\quad \quad \quad \quad \quad  a^{\dagger}_\alpha (R) \rightarrow e^{-i \theta(R)} a^{\dagger}_{\alpha}(R).  
\label{u1u1}
\eea 
Note that the above abelian gauge transformations are defined on the two sides of every link and are 
independent of the SU(2) gauge transformations (\ref{su2gt}) which are defined at every lattice site. 
Using (\ref{sb}), the electric field constraints  (\ref{E2=e2}) on the links  become the number operator 
constraints in terms of the prepotential operators: 
\bea 
\hat{N}(L) \equiv a^\dagger(L) \cdot a(L) = \hat{N}(R) \equiv a^\dagger(R) \cdot a(R) \equiv \hat{N} 
\label{noc} 
\eea 
In (\ref{noc}), $\hat{N} \equiv \hat{N}(n,i)$ and imply  $\theta(L)  = \theta(R)$ on every link and 
reduces the extra $U(1) \otimes U(1)$ gauge invariance to U(1).  Thus in the prepotential formulation 
non-abelian fluxes can be absorbed locally at a site and the abelian fluxes spread along the links. 
Both the gauge symmetries together lead to non-local (involving at least a plaquette) Wilson loop 
states (see section 3.3). 

\subsection{\bf SU(2) link operators} 

\begin{figure}[h]
\begin{center}
\includegraphics[width=0.95\textwidth,height=0.15\textwidth]
{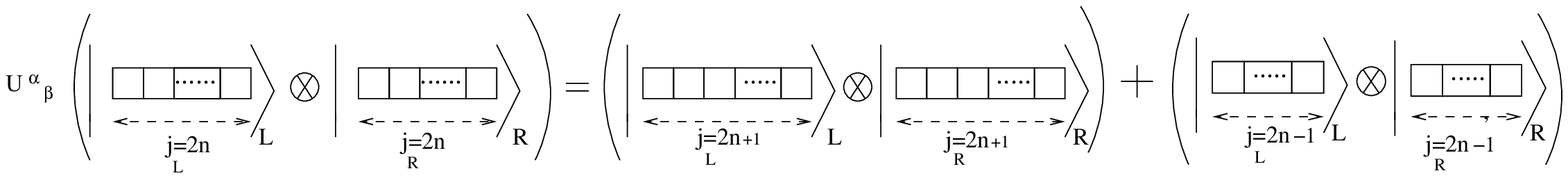}
\end{center}
\caption{The Young tableau interpretation of the SU(2) link operator U in terms of the prepotential 
operators (\ref{su2U}) acting on a state with $n_L = n_R = 2j$.
The two terms in (\ref{su2U}) correspond to the two sets of  
Young tableaues on the right hand side of this figure respectively.} 
\label{usu2}
\end{figure}

The equations (\ref{sb}) already  defines the left and right electric fields in terms of the 
prepotentials. To establish complete equivalence, we now write down the link operators explicitly 
in terms of the prepotentials. From $SU(2)$ gauge transformations of the link operator in (\ref{eugt}) 
and $SU(2) \otimes U(1)$ gauge transformations (\ref{su2gt}), (\ref{u1u1}) of the prepotentials, 
\bea
\label{su2U}
U^{\alpha}{}_{\beta}=\tilde{a}^{\dagger\alpha}(L)\, \eta \, a^{\dagger}_\beta(R) 
+ a^\alpha(L) \, \theta \, \tilde{a}_\beta(R), 
\eea
where $\eta$ and $\theta$ are functions of $SU(2)$ invariant number operator. The operators 
$\tilde{a}^{\dagger\alpha}$ and $\tilde{a}_\beta$ are defined after equation (\ref{su2gt}). 
The eqn. (\ref{su2U}) is 
graphically illustrated in terms of SU(2) Young tableaues in Figure \ref{usu2}. 
 \\
In the explicit matrix form the link operator can be written as the product of the left part $U_L$ and the 
right part $U_R$ as: 
\bea
\label{su2_U_matrix}
U=\underbrace{\left(\begin{array}{cc}
a_2^\dagger(L)\eta_L & a_1(L)\theta_L\\
-a_1^\dagger(L)\eta_L & a_2(L)\theta_L
\end{array}
\right)}_{U_L}
\underbrace{\left(\begin{array}{cc}
\eta_R a_1^\dagger(R) & \eta_R a_2^\dagger(R)\\
\theta_R a_2(R) & \theta_R (-a_1(R))
\end{array}
\right)}_{U_R}
\eea
Where, $\eta_L,\eta_R, \theta_L,\theta_R$ are the left and right invariants constructed out of number operators. 
From (\ref{su2U}) it follows that, $\eta = \eta_L\eta_R,~~~~~~\theta = \theta_L\theta_R$. 
From (\ref{su2_U_matrix}):
\bea
U_L^\dagger U_L= \left( \begin{array}{cc}
\bar{\eta}_L \left[{a^{\dagger}(L)\cdot a(L)}+2\right] \eta_L& 0\\
0 & \bar{\theta}_L  \left[a^{\dagger}(L) \cdot a(L)\right] \theta_L \end{array} \right), \nonumber \\ 
U_R U_R^\dagger= \left( \begin{array}{cc}
\eta_R \left[a^{\dagger}(R) \cdot a(R)\right] \bar{\eta}_R & 0\\
0 & {\theta_R}\left[a^{\dagger}(R) \cdot a(R) + 2 \right]\bar{\theta}_R  \end{array}
\right)
\label{ulr}
\eea
Therefore, for $U_{\alpha\beta}$  to be unitary we get: 
\bea
\eta_L = \frac{1}{\sqrt{a^{\dagger}(L)\cdot a(L)+2}},~\theta_L=\frac{1}{\sqrt{a^{\dagger}(L)\cdot a(L)}}, ~~ 
\eta_R = \frac{1}{\sqrt{a^{\dagger}(R)\cdot a(R)}},~
\theta_R=\frac{1}{\sqrt{a^{\dagger}(R)\cdot a(R)+2}}. 
\label{xyz}
\eea
Note that the operator $\eta_R$ above is always well defined as it always appears with 
$a^{\dagger}_\alpha(R)$ on it's right in (\ref{su2_U_matrix}). The operator $\theta_L$ is 
well defined in (\ref{su2_U_matrix}) 
as the link operator $U \equiv U_LU_R$ acts on the Hilbert space satisfying 
the constraints (\ref{noc}). Finally, using 
$a^{\dagger}(L) \cdot a(L) =  a^{\dagger}(R) \cdot a(R) \equiv \hat{N}$,  
the link operator can be 
disentangled  into it' left and right parts as: 
\bea
U =  \underbrace{\frac{1}{\sqrt{\hat{N}+1}}\left( \begin{array}{cc} 
a^\dagger_2({L})  & a_1(L)  \\ -a^\dagger_1(L)  & a_2(L) \end{array} 
\right) }_{U_L} \underbrace{\left( \begin{array}{cc} a^\dagger_1(R) 
& a^\dagger_2(R) \\ a_2(R) & -a_1(R)\end{array} \right) 
\frac{1}{\sqrt{\hat{N}+1}} }_{U_R} \equiv U_L ~ U_R 
\label{usu2p}
\eea 
and satisfies $U^\dagger U= UU^\dagger=1$.
\begin{figure}[t]
\begin{center}
\includegraphics[width=0.4\textwidth,height=0.4\textwidth]
{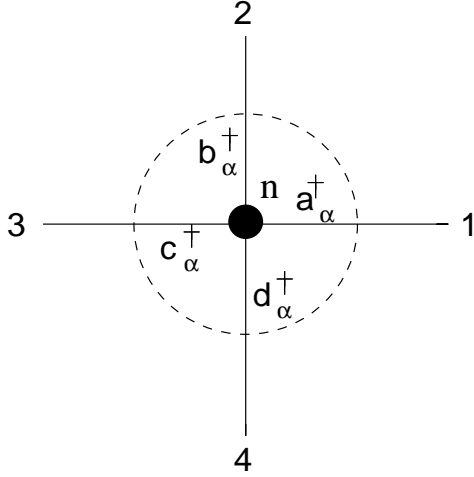}
\end{center}
\caption{SU(2) prepotentials associated with a lattice site n in a d = 2 lattice. A SU(2) gauge transformation 
at site n affects only these prepotentials enabling us to construct SU(2) gauge invariant Hilbert
spaces locally at each lattice site.}
\label{su2site}
\end{figure}

\subsection{\bf SU(2) gauge invariant states and Mandelstam constraints} 

The prepotential operators being associated with sites enable us to construct SU(2) gauge invariant 
Hilbert spaces at every lattice site. These  SU(2) gauge invariant Hilbert spaces at different lattice sites 
are mutually orthogonal. Therefore, the Mandelstam constraints which relate the various gauge invariant 
states, can be analyzed and solved locally at each lattice site.  For a $d$-dimensional lattice we have 
$2d$ number of prepotential creation operators 
present at each site   all transforming in the same way under the $SU(2)$ group present at the site (see Figure 
\ref{su2site}).  Hence all possible $SU(2)$ invariant creation operators at site $n$ are constructed by 
anti-symmetrizing any two different prepotential doublets:
\bea
\label{su2l}
L_{ij}(n)=\epsilon^{\alpha\beta}a^\dagger_\alpha(n,i)a^\dagger_\beta(n,j)=a^\dagger(n,i)\cdot \tilde
 a^\dagger(n,j),~~~~~~~~~~~~~~~~i,j=1,2,...,2d
\eea
In (\ref{su2l}), $a^\dagger_\alpha(n,i)$ with $i=1,2, \cdots 2d$ denote the 2d prepotentials around 
the lattice site n (see Figure \ref{su2site} for $d=2$). Hence, the most general gauge invariant states at a lattice site n is given by,
\bea
|\vec l(n)\rangle=\prod_{i,j=1}^{2d}\left(  L_{ij}(n) \right)^{l_{ij}(n)}|0\rangle
\eea
But these $|\vec l(n)\rangle$  states form an over complete basis because of the Mandelstam 
constraints\footnote{We will discuss the Mandelstam constraints in detail in section 4.4.} \cite{mm}: 
\bea
(a^\dagger\cdot\tilde b^\dagger)(c^\dagger\cdot\tilde d^\dagger) 
\equiv (a^\dagger\cdot\tilde c^\dagger)(b^\dagger\cdot\tilde d^\dagger)
- (a^\dagger\cdot\tilde d^\dagger)(b^\dagger\cdot\tilde c^\dagger)
\eea
A complete orthonormal gauge invariant basis at site n in terms of SU(2) prepotentials is
given in terms of SU(2) angular momentum quantum numbers \cite{mm}:   
\begin{eqnarray}
\label{std2}
|LS\rangle_{n} \equiv  |j_1,j_2,..j_{2d};j_{12},j_{123},...j_{12..(2d-1)}=j_{2d} \rangle
 =  N(j) \sum_{\{l\}}\hspace{-0.05cm}
{}^{{}^\prime} \prod_{{}^{i,j}_{i < j}}
\frac{1}{l_{ij}!}
\big({L}_{ij}(n)\big)^{l_{ij}(n)} |0 \rangle 
\end{eqnarray}
The prime over the summation means that the linking numbers $l_{ij}$ are
are summed over all possible values which are consistent with certain 
geometrical constraints \cite{mm}. The states (\ref{std2}) at different lattice sites
along with U(1) constraints (\ref{noc}) describe all possible orthonormal (linearly 
independent) loop states. It is also shown \cite{mm} that the loop dynamics for
pure SU(2) lattice gauge theory  in d dimension is given by
real and symmetric $3nj$ Wigner coefficients of the second kind 
(e.g., n=6, 10 for d=2, 3 respectively).
\begin{figure}[t]
\begin{center}
\includegraphics[width=0.8\textwidth,height=0.2\textwidth]
{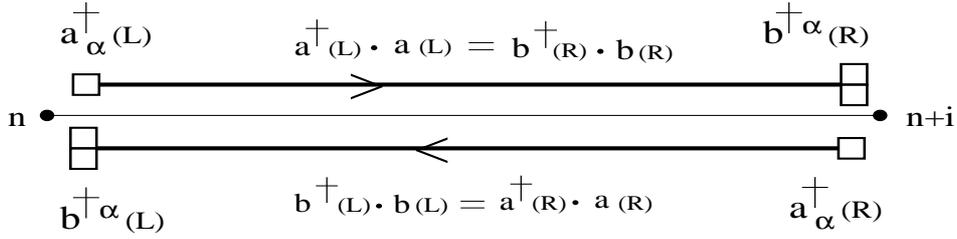}
\end{center}
\label{su3pps} 
\caption{The SU(3) prepotentials and the two $U(1) \otimes U(1)$ oriented abelian flux lines along a 
link in SU(3) lattice gauge theory. The directions of abelian flux lines are chosen from quark 
($a^{\dagger}$) prepotentials to anti-quark ($b^{\dagger}$) prepotentials.}  
\label{su3ppic}
\end{figure}
\section{Prepotentials in SU(3) lattice gauge theory}

We will now generalize the above SU(2) prepotential formulation to $SU(3)$ lattice gauge theory.
Like in SU(2), the SU(3) prepotentials are defined through the left and right electric fields 
in SU(3) lattice gauge theory. However, now  the two fundamental representations $3$ (quark) and 
$3^\ast$ (anti-quark) of $SU(3)$ are independent. Hence we associate two independent harmonic 
oscillator prepotential triplets: 
$$a^{\dagger}_{\alpha}(n,i;L) \equiv a^{\dagger}_{\alpha}(L),~~~~ 
~b^{\dagger\alpha}(n,i;L) \equiv b^{\dagger\alpha}(L), ~~~~~~\alpha=1,2,3$$  to  the left end 
and $$a^{\dagger}_{\alpha}(n+i,i;R)  \equiv a^{\dagger}_{\alpha}(R), ~~~~ 
~b^{\dagger\alpha}(n+i,i;R) \equiv b^{\dagger\alpha}(R),~~~~~~\alpha=1,2,3$$  
to the right end of the link $(n,i)$. Now there are 12 prepotential operators associated 
with every link.  These assignments are shown in Figure \ref{su3ppic}.  
Under SU(3) gauge transformation in a d dimensional spatial lattice,  the $2d$ 
$a^{\dagger}s$ and 2d  $b^{\dagger}s$ on the 2d links emanating from the lattice 
site n transform as quarks $(3)$ and anti-quarks $(3^*)$ respectively. 
The SU(3) electric fields are: 
\bea
\mbox{Left electric fields:}\quad \quad E^{\mathrm a}_L &=&  \left(a^\dagger(L) \frac{\lambda^{\mathrm a}}{2}a(L) 
- b(L)\frac{\lambda^{\mathrm a}} {2} b^\dagger(L) \right)  \nonumber \\ 
\mbox{Right electric fields:}\quad~ E^{\mathrm a}_R &=& \left(
a^\dagger(R) \frac{\lambda^{\mathrm a}}{2}a(R) - b(R)\frac{\lambda^{\mathrm a}}
{2}b^\dagger(R)\right)
\label{su3ef}
\eea
In (\ref{su3ef}), we have used Schwinger boson construction of SU(3) Lie algebra \cite{georgi,md1}. 
The electric field generators in 
(\ref{su3ef}) generate $SU_L(3) \otimes SU_R(3)$ gauge transformations on every link. 
The prepotential triplets satisfy the 
standard harmonic oscillator commutation relations: 
\bea
\Big[a^{\alpha}(l),a^\dagger_\beta(l^{\prime})\Big] & = & \delta^{\alpha}_\beta \delta_{l,l^{\prime}}~, \qquad\qquad 
\Big[b_\alpha(l),b^{\dagger\beta}(l^{\prime})\Big] = \delta_{\alpha}^\beta \delta_{l,l^{\prime}} \nonumber \\
\Big[a^{\alpha}(l),a^\beta(l^{\prime})\Big] & = & 0~, \qquad\qquad ~~~~~~
\Big[b_{\alpha}(l),b_{\beta}(l^{\prime})\Big] =  0, 
~~~~l, l^{\prime} = L,R.   
\label{comm3}
\eea
As all the electric fields in (\ref{su3ef}) involve both creation and annihilation operators, 
the number operators in (\ref{su3cas}) commute with all the electric fields in (\ref{su3ef}). 
therefore, the two SU(3) Casimirs on each side of the link (n,i) are: 
\bea 
\hat{N}(L) = a^\dagger(L) \cdot a(L), && \quad\quad\quad\quad  \hat{N}(R) = a^\dagger(R) \cdot a(R), \nonumber \\
\hat{M}(L) = b^\dagger(L) \cdot b(L), && \quad\quad\quad\quad \hat{M}(R) = b^\dagger(R) \cdot b(R).
\label{su3cas}
\eea 
The eigenvalues of $\hat{N}(L), \hat{M}(L)$ 
and $ \hat{N}(R), \hat{M}(R)$ will be denoted by $n_L, m_L$ and $n_R, m_R$ respectively. 
We can characterize all the SU(3) irreducible representations on a link by $(n_L,m_L) \otimes (n_R,m_R)$.  
Using the Gauss law generators (\ref{gl}) and the defining equations (\ref{comee}), 
the $SU(3)$ gauge transformations of the prepotentials on the left and right side of a link 
$(n,i)$ are: 
\bea 
a^{\dagger}_{\alpha}(L) \rightarrow a^{\dagger}_{\beta}(L) \big(\Lambda_L^{\dagger}\big)^{\beta}{}_{\alpha}, 
&& \quad\quad\quad\quad    
a^{\dagger}_{\alpha}(R) \rightarrow a^{\dagger}_{\beta}(R) \big(\Lambda_R^{\dagger}\big)^{\beta}{}_{\alpha} \nonumber \\ 
b^{\dagger\alpha}(L) \rightarrow \big(\Lambda_L\big)^{\alpha}{}_{\beta} b^{\dagger\beta}(L), &&   
\quad\quad\quad\quad    
b^{\dagger\alpha}(R) \rightarrow \big(\Lambda_R\big)^{\alpha}{}_{\beta} b^{\dagger \beta}(R) 
\label{su3gt} 
\eea 
The above transformations imply that under SU(3) gauge transformations 
$a^{\dagger}_{\alpha}(L), a^{\dagger}_{\alpha}(R)$ 
transform like quarks and $b^{\dagger\alpha}(L), b^{\dagger\alpha}(R)$ transform like anti-quarks at 
the left and the right end of the link $(n,i)$ respectively. 
Therefore, we call $a,a^{\dagger}$ and $b,b^{\dagger}$ on various links as 
quark and anti quark prepotentials respectively.  

\subsection{\bf The $U(1) \otimes U(1)$ gauge invariance}

Like in SU(2) case (see (\ref{u1u1})), the defining equations of SU(3) prepotentials (\ref{su3ef}) are 
invariant under the following $U(1) \otimes U(1) \otimes U(1) \otimes U(1)$ abelian gauge transformations:  
\bea 
a^{\dagger}_{\alpha}(L) \rightarrow e^{i\theta(L)} a^{\dagger}_{\alpha}(L),  && \quad\quad\quad\quad  
a^{\dagger}_{\alpha}(R) \rightarrow e^{-i \phi(R)}  a^{\dagger}_{\alpha}(R),   
\nonumber \\   
b^{\dagger\alpha}(L) \rightarrow e^{i\phi(L)} b^{\dagger\alpha}(L), &&  \quad\quad\quad\quad 
b^{\dagger\alpha} (R) \rightarrow e^{- i \theta(R)}  b^{\dagger\alpha} (R)
\label{u1u1u1}
\eea 
In (\ref{u1u1u1}), the abelian gauge angles $\theta(l)$ and $\phi(l)$ with $l=L,R$ 
are defined on the left and right sides of every link. 
Again like in SU(2) case, the Hilbert space of lattice gauge theory is built by applying 
the link operators on the vacuum state: 
$$U^{\alpha_1}{}_{\beta_1}~ U^{\alpha_2}{}_{\beta_2} \cdots U^{\alpha_n}{}_{\beta_n}|0\rangle$$ 
and then symmetrizing/anti-symmetrizing $\alpha$s  according to a certain  
Young tableau. However, this symmetrizing/anti-symmetrizing the left $\alpha \in 3$ indices 
automatically induces the same symmetries/anti-symmetries on the right $\beta \in 3^{\ast}$ indices. 
This implies that the left and right representations are always conjugate to each other\footnote{We will 
analyze the consequences of $E_L^2(n,i) = E_R^2(n+i,i)$ in the next section.}, i.e: 
\bea 
\hat{N}(L) = \hat{M}(R), ~~~~ \hat{M}(L) = \hat{N}(R).
\label{su3u1c}
\eea
This implies: $\theta(L) = \theta(R)$ and $\phi(L) = \phi(R)$ on every link. 
Therefore, besides SU(3) gauge invariance (\ref{su3gt}) at different lattice sites, 
the prepotential formulation has additional abelian $U(1) \otimes U(1)$ 
gauge invariance (\ref{u1u1u1}) on every link. 
The Gauss law constraints 
(\ref{su3u1c}) imply that abelian fluxes are oriented. We choose the directions 
of the abelian fluxes on links to be from quark to anti quark prepotentials. 
To maintain continuity of direction in a loop state the non-abelian fluxes are chosen in the opposite direction 
(i.e, from anti quark prepotentials to quark 
prepotentials).  These conventions are clearly illustrated on a link  in Figure \ref{su3ppic} and Figure \ref{k+}. 

\subsection{\bf The SU(3) prepotential Hilbert space ${\bf {\cal H}_p}$} 
Like in SU(2) case (\ref{pls}), the Hilbert space of SU(3) prepotential 
operators ${\cal H}_p$ can be completely characterized by the following 
basis on every lattice link: 
\bea 
|{}_{\alpha_1\alpha_2 \cdots \alpha_p}^{\beta_1 \beta_2 \cdots \beta_q}\rangle_L  \otimes  
|{}_{\gamma_1\gamma_2 \cdots \gamma_q}^{\delta_1 \delta_2 \cdots \delta_p}\rangle_R ~\equiv ~
\hat{L}_{~\alpha_1\alpha_2 \cdots \alpha_p}^{~\beta_1 \beta_2 \cdots \beta_q}|0\rangle_L  \otimes 
\hat{R}_{~\gamma_1\gamma_2 \cdots \gamma_q}^{~\delta_1 \delta_2 \cdots \delta_p}|0\rangle_R.  
\label{su3lst} 
\eea
In (\ref{su3lst}), 
$$\hat{L}_{\alpha_1\alpha_2 \cdots \alpha_p}^{\beta_1 \beta_2 \cdots \beta_q} |0\rangle_L 
~ \equiv ~  a^{\dagger}_{\alpha_1}(L) \cdots  a^{\dagger}_{\alpha_p}(L) b^{\dagger\beta_1}(L) \cdots  
b^{\dagger\beta_q}(L)|0\rangle_L,$$   
and 
$$\hat{R}_{\gamma_1\gamma_2 \cdots \gamma_q}^{\delta_1 \delta_2 \cdots \delta_p} |0\rangle_R 
~ \equiv ~ a^{\dagger}_{\gamma_1}(R)  \cdots  a^{\dagger}_{\gamma_q}(R)  b^{\dagger\delta_1}(R) \cdots   
b^{\dagger\delta_p}(R) |0\rangle_R$$   

\noindent are the $SU_L(3) \otimes SU_R(3) \otimes U(1) \otimes U(1)$ flux creation operators on 
the left and right ends of every link respectively.  We have used the $U(1) \otimes U(1)$ 
Gauss law constraints (\ref{su3u1c}) in (\ref{su3lst}) with $n_L = m_R =p$ and $m_L = n_R =q$.  
Note that unlike SU(2) flux creation operators (\ref{pls}) which were SU(2) irreducible, the 
flux operators operators  in (\ref{su3lst}) are SU(3) reducible (see eqns. (\ref{dps})). 
In this section we show that this is the reason why, unlike SU(2) case (\ref{hgp}), 
the SU(3) gauge theory Hilbert space ${\cal H}_g$ is contained in 
${\cal H}_p$:
\bea 
{\cal H}_g \subset {\cal H}_p. 
\label{gssp} 
\eea 
\noindent Therefore, we now need projection operators to go from ${\cal H}_p$ to ${\cal H}_g$ 
(appendix A). This makes SU(3) prepotential analysis slightly more involved than SU(2) (see 
section 4.3). To appreciate this problem, we start with 
the following SU(3) gauge invariant state 
as an example: 
\bea 
|\rho_L,~\rho_R \rangle \equiv \Big(a^{\dagger}(L) \cdot b^{\dagger}(L)\Big)^{\rho_L}  
\Big(a^{\dagger}(R) \cdot b^{\dagger}(R)\Big)^{\rho_R} | 0\rangle.   
\label{linkstate} 
\eea 
\begin{figure}[h]
\begin{center}
\includegraphics[width=1.00\textwidth,height=0.25\textwidth]
{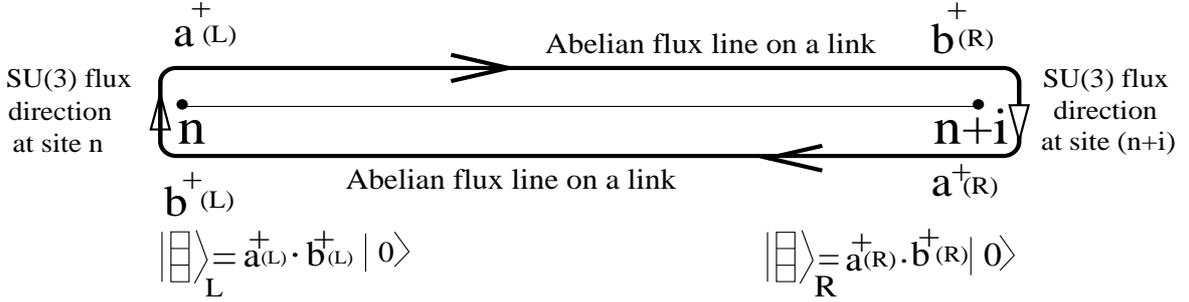}
\end{center}
\caption{The graphical interpretation of the $SU(3) \otimes U(1) \otimes U(1)$ gauge invariant 
loop state (\ref{linkstate})  over a link $(n,i)$ with $n_L=n_R=n =1$.  
The ``magnetic" Sp(2,R) quantum number $\rho$ of this state is non zero ($\rho = 1$) 
and therefore such states can not be created by the link operators $U(n,i)$. Two types 
of arrows are used to differentiate abelian and non-abelian fluxes.} 
\label{k+}
\end{figure}
The states (\ref{linkstate}) are also invariant under $U(1) \otimes U(1)$ gauge transformations 
(\ref{u1u1u1}) if $\rho_L=\rho_R =\rho$ with $\rho=0,1,2,...,\infty$. The state (\ref{linkstate}) 
with $\rho =1$ is shown in Figure \ref{k+}.  The  gauge invariant states 
(\ref{linkstate}) are linear combinations of states in (\ref{su3lst}): 
\bea 
|\rho\rangle \equiv |\rho_L =\rho,~\rho_R=\rho \rangle = \sum_{\vec{\alpha}}|{}_{\alpha_1\alpha_2 \cdots 
\alpha_{\rho}}^{\alpha_1 \alpha_2 \cdots \alpha_{\rho}}\rangle_L 
\otimes  \sum_{\vec{\beta}}|{}_{\beta_1\beta_2 \cdots \beta_{\rho}}^{\beta_1 \beta_2 \cdots \beta_{\rho}}\rangle_R. 
\label{su3ss}
\eea 
However, the only gauge invariant 
states in pure lattice gauge theories are the Wilson loop states residing around the plaquettes 
and {\it not on the links as $Tr (UU^{\dagger}) = Tr(U^{\dagger}U) = 3$} on every link. In other words, 
the infinite towers of gauge invariant states (\ref{linkstate}) 
on different links do not exist in the lattice gauge theory. 
Infact, this issue of ``non gauge theory states" in ${\cal H}_p$ is  related 
to the well known multiplicity problem in the direct products of 
SU(3). Note that the basis (\ref{su3lst}) in ${\cal H}_p$ is obtained by taking 
two direct products. 
The states $\hat{L}_{\alpha_1\alpha_2 \cdots \alpha_p}^{\beta_1 \beta_2 \cdots \beta_q} 
|0\rangle_L $  and $ \hat{R}_{\gamma_1\gamma_2 \cdots \gamma_q}^{\delta_1 \delta_2 \cdots \delta_p} 
|0\rangle_R$ are individually  direct products of quark and anti-quark irreducible representations: 
$(n_L=p,0)_L\otimes (0,m_L=q)_L$ and $(n_R=q,0)_R\otimes (0,m_R=p)_R$ respectively. Therefore, they 
can be further reduced using the SU(3) Clebsch Gordan series into irreps. of $SU_L(3)$ and $SU_R(3)$ 
respectively: 
\bea 
(n_L=p,0)_L \otimes (0,m_L=q)_L & = & \sum_{\rho(L)=0}^{min(p,q)} \oplus 
\underbrace{(p -\rho(L),q-\rho(L))_L}_{{\cal H}_p^L(p-\rho(L),q-\rho(L),\rho(L))}, \nonumber \\ 
(n_R=q,0)_R \otimes (0,m_R=p)_R & = & \sum_{\rho(R)=0}^{min(p,q)} \oplus 
\underbrace{(q -\rho(R),p-\rho(R))_R}_{{\cal H}_p^R(q-\rho(R),p-\rho(R),\rho(R))}.  
\label{dps} 
\eea
The  multiplicities\footnote{Under SU(3) gauge transformations, the vectors in ${\cal H}_p^{L}(p,q,\rho) \otimes 
{\cal H}_p^{R}(q,p,\rho)$ in (\ref{cpphs}) transform as $(p,q)_L \otimes (q,p)_R$ irreducible representation of 
$SU_L(3) \otimes SU_R(3)$  independent of the value of $\rho~ (=0,1, \cdots,
\infty)$ leading to infinite multiplicity for each state. While the gauge theory Hilbert space ${\cal H}_g$ 
contains each of these representations only once (see (\ref{ginp})).} occurring in such direct product representations 
have been extensively studied and classified in \cite{nm}. 
Following \cite{nm}, we have defined ${\cal H}^l_p(p-\rho(l),q-\rho(l),\rho(l)), 
l =L,R$ mutually orthogonal Hilbert spaces as these Hilbert spaces are in different irreducible 
representations of $SU_l(3)$. As shown in appendix B, the SU(3) electric field constraints 
$E^2_L(n,i) = E^2_R(n+i,i)$ along with the $U(1) \otimes U(1)$ Gauss law constraints (\ref{su3u1c}) 
on links imply: 
$$\rho(n,i;L) = \rho(n+i,i;R)$$ 
in the SU(3) Clebsch Gordan series (\ref{dps}) on every 
link. Therefore, the prepotential Hilbert space can be classified as:  
\bea 
{\cal H}_p  = \prod_{\otimes link} \Big\{{\cal H}_p\Big\}_{link}  & = & 
\prod_{\otimes link} \Big\{\sum_{\rho=0}^{\infty} ~\sum_{p,q=0}^{\infty}\Big({\cal H}_p^L(p,q,\rho) 
\otimes {\cal H}_p^R(q,p,\rho)\Big)\Big\}_{{\textrm link}}  \nonumber \\
& \equiv & \prod_{\otimes link} \Big\{\sum_{\rho=0}^{\infty}{\cal H}_p(\rho)\Big\}_{link}. 
\label{cpphs} 
\eea 
In order to identify the gauge theory Hilbert space ${\cal H}_g$ in (\ref{cpphs}), we define the 
following three color neutral operators on each side $l$ of every link: 
\bea 
k_{-}(l) \equiv a(l) \cdot b(l), ~k_{+}(l) \equiv a^{\dagger}(l)\cdot b^{\dagger}(l), ~k_{0}(l) \equiv \frac{1}{2} 
\left(a^{\dagger}(l) \cdot a(l) + b^{\dagger}(l) \cdot b(l) + 3\right), ~~ l \equiv L,R.   
\label{sp2rg} 
\eea 
As usual, we have suppressed the link indices $(n,i)$ in (\ref{sp2rg}). 
These SU(3) color neutral operators satisfy the Sp(2,R) algebra on both sides 
of the link:
\bea
[k_0(l),k_{\pm}(l^{\prime})]=\pm \delta_{l,l^{\prime}} k_{\pm}(l),\qquad\qquad 
[k_-(l),k_+(l^{\prime})]=2 \delta_{l,l^{\prime}} k_0(l), ~~ l,l^{\prime} \equiv L,R. 
\label{sp2r}
\eea
Further, as these Sp(2,R) generators are invariant under SU(3) transformations, 
they  commute with the color electric fields. In other words: 
\bea 
\left[Sp_L(2,R) \otimes Sp_R(2,R), SU_L(3) \otimes SU_R(3)\right] = 0.  
\label{cnc}
\eea 
Therefore, the Hilbert space of SU(3) lattice gauge theory can be completely and uniquely labeled by 
$SU_L(3) \otimes  Sp_L(2,R) \otimes SU_R(3) \otimes Sp_R(2,R)$  quantum numbers on every link.  
The irreducible representations of Sp(2,R) are characterized by $|k,\rho\rangle$, where 
$k$ and $\rho$ represent the Sp(2,R) ``spin" and ``magnetic" quantum numbers. 
For the direct product (\ref{dps}) we get \cite{nm}: $k(L) = k(R) = \frac{1}{2}(p+q+3)$. 
Further  $\rho(L) = \rho(R)$ appearing in  (\ref{dps}) are the 
``magnetic quantum numbers" of  $Sp_L(2,R) \otimes Sp_R(2,R)$. The raising (lowering) 
$K_+ (K_-)$ operators increase (decrease) the Sp(2,R) magnetic fluxes \cite{nm}: 
\bea 
|{\cal H}^L_{p}(p,q,\rho \pm 1)\rangle   & = &  k_{\pm}(L)~ |{\cal H}^L_{p}(p,q,\rho)\rangle, \quad \quad      
|{\cal H}^R_{p}(q,p,\rho \pm 1)\rangle   =  k_{\pm}(R)~ |{\cal H}^R_{p}(q,p,\rho)\rangle, 
\label{idsp2rm} 
\eea
where $|{\cal H}^l_{p}(p,q,\rho)\rangle$ denotes an arbitrary vector in ${\cal H}^l_{p}(p,q,\rho)$ with 
$l=L/R$.  In particular, the $\rho = 0$ Hilbert space without any ``Sp(2,R) magnetic" flux in (\ref{dps}) 
is annihilated by $k_-$: 
\bea 
k_-(L)~|{\cal H}^L_{p}(p,q,\rho=0)\rangle   = 0, \quad \quad \quad \quad \quad \quad  \quad \quad 
k_-(R)~|{\cal H}^R_{p}(q,p,\rho=0)\rangle   =   0.    
\label{kan} 
\eea 
The equations (\ref{idsp2rm}) show that the ``spurious gauge invariant" states in  (\ref{linkstate})  
are the vectors of one dimensional mutually orthogonal SU(3) invariant Hilbert spaces ${\cal H}_p^L(0,0,\rho) 
\otimes {\cal H}_p^R(0,0,\rho)$ with $\rho =1, \cdots \infty$.  The strong coupling vacuum is the $\rho =0$ vacuum. 

\subsection{The SU(3) gauge theory Hilbert space ${\cal H}_g$} 

The various flux states in gauge theory Hilbert space ${\cal H}_g$ are created by the link 
matrices $U^\alpha{}_\beta$ acting on the strong coupling vacuum as in (\ref{feq}). Therefore, 
in order to identify ${\cal H}_g$ in ${\cal H}_p$ with Sp(2,R) structure (\ref{cpphs}), we now analyze the
Sp(2,R) properties of the link operators in this section. We note that the link matrix $U^{\alpha}{}_\beta$ 
can not change the Sp(2,R) magnetic quantum number $\rho$. 
As shown at the bottom of Figure \ref{k+},  $k_+(L) = a^{\dagger}(L) \cdot b^{\dagger}(L)$ and 
$k_+(R) = a^{\dagger}(R) \cdot b^{\dagger}(R)$ correspond to 
three Young tableau boxes in a vertical 
column (SU(3) singlets)  on the left and right side of the links respectively. On the other hand, 
in terms of the link operators, 
this left and right anti-symmetrization on a link corresponds to: $\frac{1}{3!} \epsilon_{\alpha_1\alpha_2\alpha_3} 
\epsilon^{\beta_1\beta_2\beta_3} U^{\alpha_1}{}_{\beta_1} U^{\alpha_2}{}_{\beta_2} U^{\alpha_3}{}_{\beta_3} = 
{\textrm det}~ U \equiv 1$ or $tr~(U U^{\dagger})=3$. Therefore,  the states in 
${\cal H}_g$, obtained by applying link operators on the strong coupling 
vacuum with $\rho=0$ ($k_-(l)|0\rangle_l =0, ~ l= L,R$) will also carry $\rho =0$ quantum numbers. 
In other words, they too will be annihilated by $k_-(l)$: 
\bea 
k_-(L)\Big( U^{\alpha_1}{}_{\beta_1} U^{\alpha_2}{}_{\beta_2} \cdots U^{\alpha_r}{}_{\beta_r}\Big)|0\rangle  =  0,~~ 
\quad \quad k_-(R)\Big( U^{\alpha_1}{}_{\beta_1} U^{\alpha_2}{}_{\beta_2} \cdots U^{\alpha_r}{}_{\beta_r}\Big)|0\rangle  =  0. 
\label{kmcu} 
\eea    
Therefore, going back to the classification of ${\cal H}_p$ in (\ref{cpphs}), we identify: 
\bea 
{\cal H}_g  \equiv \prod_{\otimes link} \big\{{\cal H}_p(\rho =0)\big\}_{link} \equiv {\cal H}_p^0 
\label{ginp}
\eea 
like in the case of SU(2) lattice gauge theory. 
In (\ref{ginp}) ${\cal H}_p^0$ denotes $\rho =0$ subspace of ${\cal H}_p$. 
Thus the kernel of $\left(k_-(L) k_-(R)\right)$ in ${\cal H}_p$ is the SU(3) gauge theory 
Hilbert space ${\cal H}_g$. Further, (\ref{kmcu}) implies: 
\bea 
\left[k_-(L),U^{\alpha}{}_\beta\right] \simeq 0, \quad \quad\quad\quad \left[k_-(R),U^{\alpha}{}_\beta\right] \simeq 0, 
\label{sp2rc}
\eea
In other words, $k_-(L)$ and $k_-(R)$ weakly commute with the link operators of SU(3) lattice gauge 
theory\footnote{Note that all the electric fields strongly commute with the Sp(2,R) generators 
(\ref{cnc}).}. The symbol $\simeq$ in (\ref{sp2rc}) implies that the commutators are zero only 
when they are applied on the vectors belonging to the gauge theory Hilbert space ${\cal H}_g$. 
We would now like to write the link operators in terms of SU(3) prepotential operators 
which create SU(3) fluxes only in the gauge theory Hilbert space ${\cal H}_g$. This is done in 
the next section.  

\subsection{\bf SU(3) irreducible prepotential operators} 

In this section, we construct the SU(3) irreducible prepotential operators 
from the  prepotential operators in (\ref{su3ef}) such that they directly create SU(3) irreducible 
fluxes exactly like in SU(2) case (\ref{pls}). This construction with all the it's group theoretical 
details is given in \cite{ima}.  We define the SU(3) irreducible 
prepotential operators from prepotential operators such that: 
\begin{enumerate} 
\item  they have exactly the same $SU(3) \otimes U(1) \otimes U(1)$ quantum numbers, 
\item  they commute with the Sp(2,R) destruction operator $k_-$. 
\end{enumerate} 
As a result, acting on the strong coupling 
vacuum they directly create the gauge theory Hilbert space ${\cal H}_g$ completely 
bypassing the problem of spurious states like (\ref{linkstate}) in ${\cal H}_p$. 
we define SU(3) irreducible prepotentials \cite{ima} as: 
\bea
\label{modsb2}
{A}^\dagger_\alpha(L)   & = &  a^\dagger_\alpha(L)- F_L ~  k_+(L)b_\alpha(L), ~~~~~~
{A}^\dagger_\alpha(R)  =  a^\dagger_\alpha(R)- F_R ~  k_+(R)b_\alpha(R), \nonumber \\ 
{B}^{\dagger\alpha}(L) &  = &  b^{\dagger\alpha}(L)-  F_L k_+(L) a^\alpha(L), ~~~~~~
{B}^{\dagger\alpha}(R)  =  b^{\dagger\alpha}(R)-  F_L k_+(R) a^\alpha(R). 
\eea
In (\ref{modsb2}), the factors $F_L$ and $F_R$ are given by: 
$$F_L = \frac{1}{N(L) + M(L) + 1},~~~   F_R = \frac{1}{N(R) + M(R) + 1}.$$  
These factors are chosen so that \cite{ima}: 
\bea 
\left[k_-(l), {A}^\dagger_\alpha(l)\right] \simeq 0; \quad \quad \quad \quad   
~~~~ \left[k_-(l), {B}^{\dagger\alpha}(l)\right] \simeq 0. 
\label{all0} 
\eea 
It is easy to check that the irreducible Schwinger boson creation operators
commute amongst themselves:
\bea
\left[A^{\dagger}_{ \alpha}(l), A^{\dagger}_{ \beta}(l^\prime)\right] = 0, ~~~~ \left[B^{\dagger \alpha}(l),
B^{\dagger \beta}(l^\prime)\right] = 0, ~~~~ \left[A^{\dagger}_{\alpha}(l), B^{\dagger \beta}(l^\prime)\right] =0.
\label{comm}
\eea
The other commutation relations acting on the SU(3) irreps. are \cite{ima}:
\bea
\left[A^{\alpha}(l), A^{\dagger}_{\beta}(l^\prime)\right]
& \simeq &
\delta_{l l^\prime} 
\left(\delta_{\alpha}^{\beta} - \frac{1}{N(l)+M(l)+2} B^{\dagger\alpha}B_{\beta}\right)
\nonumber \\
\label{commr}
\left[A^{\alpha}(l), B^{\dagger\beta}(l^\prime)\right]
& \simeq & 
~~ -\delta_{l l^\prime} 
\frac{1}{N(l)+M(l)+2} B^{\dagger\alpha} A^{\beta}
\\
\left[B_{\alpha}(l), B^{\dagger\beta}(l^\prime)\right]
& \simeq &
\delta_{l l^\prime} 
\left(\delta^{\alpha}_{\beta} - \frac{1}{N(l)+M(l)+2} A^{\dagger}_{\alpha}A^{\beta}\right)
\nonumber
\eea
By construction, ${A}^\dagger_\alpha(l)$ and ${B}^{\dagger\alpha}(l)$ transform exactly 
like ${a}^\dagger_\alpha(l)$ and ${b}^{\dagger\alpha}(l)$, $l = L,R$ under $SU(3) \otimes U(1) 
\otimes U(1)$ and retain the same quantum numbers. Therefore, we can now define: 
\bea 
|{}_{\alpha_1\alpha_2 \cdots \alpha_p}^{\beta_1 \beta_2 \cdots \beta_q}\rangle_L^{0}   \otimes  
|{}_{\gamma_1\gamma_2 \cdots \gamma_q}^{\delta_1 \delta_2 \cdots \delta_p}\rangle_R^{0} ~\equiv ~
\hat{{\cal L}}_{~\alpha_1\alpha_2 \cdots \alpha_p}^{~\beta_1 \beta_2 \cdots \beta_q}|0\rangle_L  \otimes 
\hat{{\cal R}}_{~\gamma_1\gamma_2 \cdots \gamma_q}^{~\delta_1 \delta_2 \cdots \delta_p}|0\rangle_R.  
\label{su3lstir} 
\eea
In (\ref{su3lstir}), the additional Sp(2,R) quantum numbers $\rho_L = \rho_R =0$ are put as 
superscript 0. The operators ${\cal L}$ and ${\cal R}$ are defined by replacing SU(3) prepotentials 
in L and R in (\ref{su3lst}) by the corresponding SU(3) irreducible prepotentials in (\ref{modsb2}), i.e.,    
$$\hat{{\cal L}}_{\alpha_1\alpha_2 \cdots \alpha_p}^{\beta_1 \beta_2 \cdots \beta_q} |0\rangle_L 
~ \equiv ~  A^{\dagger}_{\alpha_1}(L) \cdots  A^{\dagger}_{\alpha_p}(L) B^{\dagger\beta_1}(L) \cdots  
B^{\dagger\beta_q}(L)|0\rangle_L,$$   
and 
$$\hat{{\cal R}}_{\gamma_1\gamma_2 \cdots \gamma_q}^{\delta_1 \delta_2 \cdots \delta_p} |0\rangle_R 
~ \equiv ~ A^{\dagger}_{\gamma_1}(R)  \cdots  A^{\dagger}_{\gamma_q}(R)  B^{\dagger\delta_1}(R) \cdots   
B^{\dagger\delta_p}(R) |0\rangle_R.$$   
Note that in terms of SU(3) irreducible prepotentials, the ``spurious gauge invariant states" like in 
(\ref{linkstate}) or (\ref{su3ss}) do not exist as: 
\bea 
A^{\dagger}(L) \cdot B^{\dagger}(L)|0\rangle_L \equiv  0, \quad \quad\quad\quad\quad\quad 
A^{\dagger}(R) \cdot B^{\dagger}(R)|0\rangle_R \equiv  0. 
\eea 
In other words, the operators ${\cal L}$ and ${\cal R}$ in (\ref{su3lstir}) are SU(3) irreducible unlike 
the L and R operators in (\ref{su3lst}) which are reducible according to (\ref{dps}). In the appendix A 
we show that ${\cal L}$  and ${\cal R}$ are related to L and R by  projection operators (\ref{cptp}). 
Infact, these 
SU(3) flux creation operators ${\cal L}$ and ${\cal R}$  are the SU(3) analogues of 
the SU(2) flux creation operators ${\cal L}$ and ${\cal R}$ in (\ref{pls}) as both create 
irreducible fluxes. Further, like in SU(2) case, they bypass the problem of 
symmetrization and anti symmetrization associated with the link operators. 
This is because $\hat{\cal L}$ and $\hat{\cal R}$ in (\ref{su3lstir})  are defined in terms of 
SU(3) irreducible prepotential operators which have  all the symmetries of SU(3) Young tableaues 
inbuilt \cite{ima}.  In other words the role played by SU(2)  prepotentials in SU(2) lattice gauge theory is exactly  
equivalent to the role played by SU(3) irreducible prepotentials in SU(3) lattice gauge theory. 

\subsection{\bf SU(3) link operators} 

The SU(3) link operator must create $3$ and $3^*$ fluxes 
at the left and right end of the link and should satisfy the $U(1) \otimes U(1)$ Gauss law constraints 
(\ref{su3u1c}). These requirements are similar to SU(2) case discussed in section 3.2. The new 
SU(3) requirement of Sp(2,R) constraint (\ref{sp2rc}) has been solved by defining SU(3) irreducible 
prepotential operators in the previous section. 
Noting that by construction, ${A}^\dagger_\alpha(l)$ and ${B}^{\dagger\alpha}(l)$ transform exactly 
like ${a}^\dagger_\alpha(l)$ and ${b}^{\dagger\alpha}(l)$, $l = L,R$,  the general structure of the 
link operator is: 
\begin{figure}[t]
\begin{center}
\includegraphics[width=0.85\textwidth,height=0.12\textwidth]
{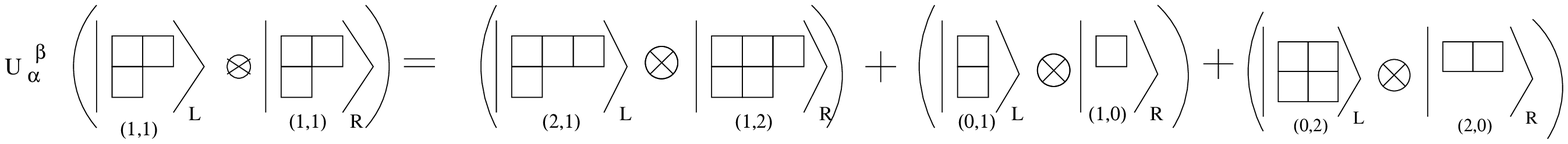}
\end{center}
\caption{The Young tableau interpretation of the SU(3) link operator U in terms of the prepotential 
operators (\ref{su3U}) acting on a state with $n_L = m_R \equiv p = 1$ and $m_L = n_R \equiv q =1$.
The three terms in (\ref{su3U}) or (\ref{cgcs})  correspond to the three sets of (mutually conjugate) 
Young tableaues on the right hand side of this figure respectively. This is SU(3) generalization of 
Figure \ref{usu2} for SU(2).}    
\label{su3ytm}
\end{figure}
\bea
U^\alpha{}_\beta &=&  B^{\dagger\alpha}(L) ~{\eta} ~A^{\dagger}_{\beta}(R) +
A^\alpha(L)~  {\theta}~ B_\beta(R) + \Big(B(L) \wedge {A^\dagger}(L)\Big)^{\alpha}~ {\delta}~ 
\Big(A(R) \wedge B^\dagger(R)\Big)_{\beta}. 
\label{su3U}
\eea
In (\ref{su3U}), $\eta,\theta$ and $\delta$ are the SU(3) invariants and therefore can only  depend 
on the number operators. These will be fixed later in this section. 
The link operator constructed in (\ref{su3U}) has all the required group theoretical properties:
\begin{itemize} 
\item Under SU(3) transformations $U(n,i)^{\alpha}{}_{\beta}  \rightarrow 
(\Lambda_L)^{\alpha}{}_{\gamma} U(n,i)^{\gamma}{}_{\delta}({\Lambda_R}^{\dagger})^{\delta}{}_{\beta}$. 
\item It is invariant under  $U(1) \otimes U(1)$ abelian gauge transformations. 
\item  It creates and destroys fluxes in ${\cal H}_p^0$ in (\ref{ginp}). It is easy to check that the link 
operator $U^\alpha{}_\beta$ in (\ref{su3U}) satisfy (\ref{sp2rc}). 
\item Acting on a link state in  $(p,q)_L$ and $(q,p)_R$ representations of $SU(3)_L\times SU(3)_R$:
\bea 
U^{\alpha}{}_{\beta} |p,q\rangle_L \otimes |q,p\rangle_R  & = & 
C_1{}^{\alpha}{}_{\beta} |p+1,q\rangle_L \otimes |q,p+1\rangle_R 
 +  
C_2{}^{\alpha}{}_{\beta} |p,q-1\rangle_L \otimes |q-1,p\rangle_R \nonumber \\
&+ &C_3{}^{\alpha}{}_{\beta} |p-1,q+1\rangle_L \otimes |q+1,p-1\rangle_R,
\label{cgcs} 
\eea 
where $C_1, C_2$ and $C_3$ are the SU(3) Clebsch Gordan coefficients. The three terms in (\ref{su3U}) 
correspond to the three terms in (\ref{cgcs}) respectively. In Figure \ref{su3ytm}, we illustrate 
(\ref{su3U}) and (\ref{cgcs}) in terms of SU(3) Young tableau diagrams. 
\end{itemize}  
\noindent Like in SU(2) case, it is convenient to define left and right link operators as: 
\bea
U=
\underbrace{\left( \begin{array}{ccc} B^{\dagger 1}(L) \eta_{L} & A^1(L) \theta_{L} & (B(L) \wedge A^\dagger(L))^1 
\delta_{L}\\ 
B^{\dagger 2}(L) \eta_{L} & A^2(L) \theta_{L} & (B(L) \wedge A^\dagger(L))^2 \delta_{L} \\
B^{\dagger 3}(L) \eta_{L} & A^3(L) \theta_{L} & (B(L) \wedge A^\dagger(L))^3 \delta_{L}\end{array} \right) }_{U_L} 
\underbrace{\left( \begin{array}{ccc} A^1(R) 
\bar\eta_{R} & B^{1\dagger}(R) \bar\theta_{R}  & (B(R) \wedge A^\dagger(R))^1 
\bar\delta_{R}\\ 
A^2(R) \bar\eta_{R} & B^{2 \dagger}(R) \bar\theta_{R} & (B(R) \wedge A^\dagger(R))^2 \bar\delta_{R}  \\
A^3(R) \bar\eta_{R} & B^{3 \dagger}(R) \bar\theta_{R} & (B(R) \wedge A^\dagger(R))^3 \bar\delta_{R}\end{array} 
\right)^{\dagger} }_{U_R} 
\label{lru} 
\eea
Where $\eta_L, \theta_L,  \delta_L$ and $ \bar\eta_R, \bar\theta_R,  \bar\delta_R$ are the  left  and  right invariants 
constructed out of the number operators.  From (\ref{su3U}): 
\bea  
\eta = \eta_L ~ {\eta}_R,~~~~~~  \theta = \theta_L ~ {\theta}_R,~~~~~~  \delta = \delta_L ~ {\delta}_R.  
\label{cno}
\eea
From (\ref{lru}): 
\bea 
U_L^{\dagger} U_L =  
\left( \begin{array}{ccc} \bar\eta_L\Big( B \cdot B^{\dagger}\Big) \eta_{L} 
& \bar\eta_L \underbrace{\Big( B \cdot A\Big)}_{\simeq 0}  \theta_L & \bar\eta_L 
\underbrace{\Big(B \cdot \left(B \wedge A^\dagger \right) \Big)}_{\equiv 0}  \delta_L \\ 
\bar\theta_L \underbrace{\Big(A^{\dagger} \cdot B^\dagger\Big)}_{\simeq 0}  \eta_{L}  
& \bar\theta_L \Big( A^\dagger \cdot A
\Big)  \theta_{L} & \bar\theta_L \underbrace{\Big( A^{\dagger} \cdot \left(B \wedge A^\dagger\right) 
\Big)}_{\equiv 0} \delta_{L} \\
\bar\delta_L \underbrace{\Big(\left(B^{\dagger} \wedge A\right) \cdot B^{\dagger} \Big)}_{\equiv 0} \eta_{L}  
& \bar\delta_L \underbrace{\Big(\left(B^{\dagger} \wedge A \right) \cdot A \Big)}_{\equiv 0} \theta_{L}  
& \bar\delta_L \Big(\left(A\wedge B^\dagger \right) \cdot  \left(B \wedge A^\dagger\right)\Big) \delta_{L}\end{array} \right)
\label{luc} 
\eea 
Similarly, 
\bea 
U_R U^\dagger_R =  
\left( \begin{array}{ccc} \eta_R\Big( A^\dagger \cdot A\Big) \bar\eta_{R} 
& \eta_R \underbrace{\Big( A^\dagger \cdot B^\dagger \Big)}_{\simeq 0} \bar\theta_R & \eta_R 
\underbrace{\Big(A^\dagger \cdot \left(B \wedge A^\dagger \right) \Big)}_{\equiv 0} \bar\delta_R \\ 
\theta_R \underbrace{\Big(B \cdot A\Big)}_{\simeq 0}  \bar\eta_{R}  & \theta_R \Big( B \cdot B^\dagger 
\Big)  \bar\theta_{R} & \theta_R \underbrace{\Big( B \cdot \left(B \wedge A^\dagger\right) 
\Big)}_{\equiv 0} \bar\delta_{R} \\
\delta_R \underbrace{\Big(\left(B^{\dagger} \wedge A\right) \cdot A \Big)}_{\equiv 0}  \bar\eta_{R}  
& \delta_R \underbrace{\Big(\left(B^{\dagger} \wedge A \right) \cdot B^\dagger \Big)}_{\equiv 0}\bar\theta_{R}  
& \delta_R \Big(\left(A\wedge B^\dagger \right) \cdot  \left(B \wedge A^\dagger \right)\Big)
 \bar\delta_{R}\end{array} \right)
\label{ruc} 
\eea 
In (\ref{luc}) and (\ref{ruc}), we have suppressed the L/R indices from the prepotential operators $(A,A^{\dagger})$ 
and $(B, B^{\dagger})$.  Demanding $U_L^\dagger U_L =1$ and $U_RU_R^\dagger=1$, we get: 
\bea 
\eta_L = \frac{1}{\sqrt{B(L) \cdot B^{\dagger}(L)}}, ~~\theta_L = \frac{1}{\sqrt{A^{\dagger}(L) \cdot A(L)}},~~ 
\delta_L =  \frac{1}{\sqrt{\left(A(L)\wedge B^\dagger(L) \right) \cdot  \left(B(L) \wedge A^\dagger(L)\right)}}; \nonumber \\
\eta_R = \frac{1}{\sqrt{A^\dagger (R) \cdot A(R)}}, ~~
\theta_R = \frac{1}{\sqrt{B(R) \cdot B^\dagger(R)}},~~ 
\delta_R =  \frac{1}{\sqrt{\left(A(R)\wedge B^\dagger(R) \right)\cdot \left(B(R) \wedge A^\dagger(R)\right)}}. 
\label{coeff} 
\eea
The link operators in (\ref{su3U}) with (\ref{cno}) and (\ref{coeff}) satisfy: $U U^{\dagger} = U^\dagger U =1$.  
Having written the link operators in terms of the SU(3) irreducible prepotentials, we now cast the left and right 
electric fields  (\ref{su3ef}) in terms of $A(l), A^{\dagger}(l), B(l),  B^{\dagger}(l)$ with $l=L,R$. Using the 
very special structures of the SU(3) irreducible prepotentials in (\ref{modsb2}), it is easy to check that:  
\bea
E^{\mathrm a}_L &=& 
 \left(a^\dagger(L) \frac{\lambda^{\mathrm a}}{2}a(L) - b(L)\frac{\lambda^{\mathrm a}} {2} b^\dagger(L) \right)  
\simeq 
\left(A^\dagger(L) \frac{\lambda^{\mathrm a}}{2}A(L) - B(L)\frac{\lambda^{\mathrm a}} {2} B^\dagger(L) \right)  
\nonumber \\ 
E^{\mathrm a}_R &=& 
\left(a^\dagger(R) \frac{\lambda^{\mathrm a}}{2}a(R) - b(R)\frac{\lambda^{\mathrm a}} {2}  b^\dagger(R) \right)    
\simeq 
\left(A^\dagger(R) \frac{\lambda^{\mathrm a}}{2}A(R) - B(R)\frac{\lambda^{\mathrm a}} {2}  B^\dagger(R)\right)
\label{su3efn}
\eea
In (\ref{su3efn}), we have made use of the identities: $a(L) \cdot b(L) \equiv k_-(L) \simeq 0$ and 
$a(R) \cdot b(R) \equiv k_-(R) \simeq 0$ on every link of the lattice. In fact the results (\ref{su3efn}) 
were expected because  $(a^{\dagger}_{\alpha}, b^{\dagger\beta})$ and $(A^{\dagger}_{\alpha}, B^{\dagger\beta})$ 
have exactly the same $SU(3) \otimes U(1) \otimes U(1)$ transformation properties. 
\begin{figure}[h]
\begin{center}
\includegraphics[width=0.4\textwidth,height=0.4\textwidth]
{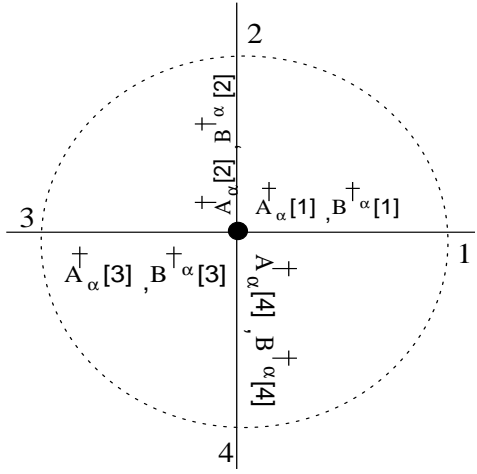}
\end{center}
\caption{The SU(3) prepotentials associated with a lattice site n in $d=2$. This is SU(3) generalization of 
Figure (\ref{su2site}) for SU(2). } 
\label{su3s}
\end{figure}

\subsection{SU(3) gauge invariant states and Mandelstam Constraints}

\begin{figure}[t]
\begin{center}
\includegraphics[width=0.6\textwidth,height=0.31\textwidth]
{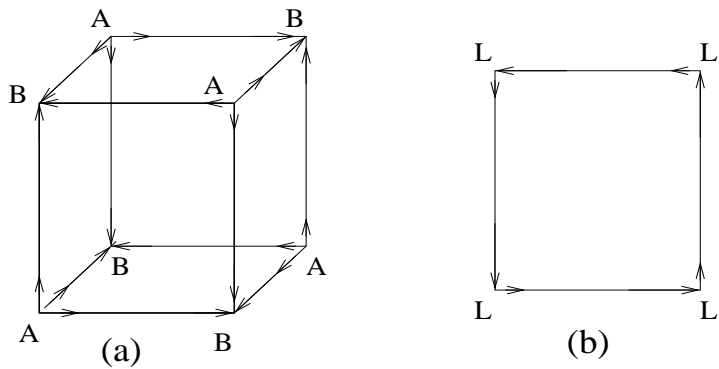}
\end{center}
\caption{Graphical representation of the three possible SU(3) gauge invariant L, A, B types of vertices. 
Two simple $SU(3) \otimes U(1) \otimes U(1)$  gauge invariant loop states are also shown.  
The arrows represent the directions of the abelian (non-abelian) fluxes on the links (sites).} 
\label{su3gisp}
\end{figure}

In this section we construct all possible SU(3) gauge invariant states at 
a given lattice site using prepotential approach. We also discuss the Mandelstam constraints 
which relate these gauge invariant states. 
The additional $U(1) \otimes U(1)$ Gauss law (\ref{su3u1c})  can be satisfied by drawing the abelian 
flux lines  along the links as is done in  Figure \ref{su3gisp}. 
As shown in Figure \ref{su3s}, every lattice site in 2d space dimension is associated with $2d$ pairs of 
quark-anti quark prepotentials $(A^{\dagger}_{\alpha},B^{\dagger\alpha})$. Under a gauge transformation at site n, 
all these $2d$ quark (anti quark) prepotentials transform together as triplet (anti-triplet). Therefore, 
the fundamental SU(3) gauge invariant creation operator vertices at a lattice site n are: 
\bea
\label{gi1} 
L_{[ij]}& \equiv & A^{\dagger}[i]\cdot B^\dagger[j], ~~~~~~~~~~~  i\neq j,\\ 
\label{gi2} 
A_{[i_1,i_2,i_3]} &=&\epsilon^{\alpha_1\alpha_2\alpha_3}
A^{\dagger}_{\alpha}[i_1]A^{\dagger}_{\alpha_2}[i_2]A^{\dagger}_{\alpha_3}[i_3], \\
\label{gi3} 
B_{[j_1,j_2,j_3]} &=& \epsilon_{\beta_1\beta_2\beta_3}  
B^{\dagger\beta_1}[j_1]B^{\dagger\beta_2}[j_2]B^{\dagger\beta_3}[j_3]
\eea 
These vertices are shown in Figure \ref{su3gisp}. 
We have taken $i\neq j$ in (\ref{gi1}) because $L_{ii} = A^{\dagger}[i]\cdot B^\dagger[i] \simeq 0, ~~i,j=1,2, \cdots
2d$ according to (\ref{sp2rc}).  Also, $A_{[i_1,i_2,i_3]}$ and $B_{[j_1,j_2,j_3]}$ are completely anti-symmetric in 
$(i_1,i_2,i_3)$ and $(j_1,j_2,j_3)$ indices respectively. The above $$2({}^{2d}C_2) + 2({}^{2d}C_3) = 
\frac{2d(2d-1)(2d+1)}{3}$$ basic SU(3) gauge invariant operators  enable us  to write the most general SU(3) 
gauge invariant state at a given lattice site as: 
\bea 
\vert \vec{l}_{[ij]}, \vec{p}_{[i_1i_2i_3]}, \vec{q}_{[j_1j_2j_3]} \rangle = \prod_{{}^{i,j=1}_{~i\neq j}}^{2d} 
\Big(L_{[ij]}\Big)^{l_{[ij]}} 
\prod_{[i_1i_2i_3]=1}^{{}^{2d}C_3} \Big(A_{[i_1i_2i_3]}\Big)^{p_{[i_1i_2i_3]}}
\prod_{[j_1j_2j_3]=1}^{{}^{2d}C_3} \Big(B_{[j_1j_2j_3]}\Big)^{q_{[j_ji_2j_3]}} |0\rangle.  
\label{su3gis}
\eea    

\noindent In (\ref{su3gis}), $\vec{l}_{[ij]}, \vec{p}_{[i_1i_2i_3]}, \vec{q}_{[j_1j_2j_3]}$ are $\frac{2d(2d-1)(2d+1)}{3}$
non-negative integers describing all possible SU(3) gauge invariant states at a given lattice site. The 
various possible loop states set in pure SU(3) lattice gauge theory are direct products of (\ref{su3gis}) 
at various lattice sites consistent with $U(1) \otimes U(1)$ Gauss law (\ref{su3u1c}) along every link.  

\begin{figure}[t]
\begin{center}
\includegraphics[width=0.9\textwidth,height=0.25\textwidth]
{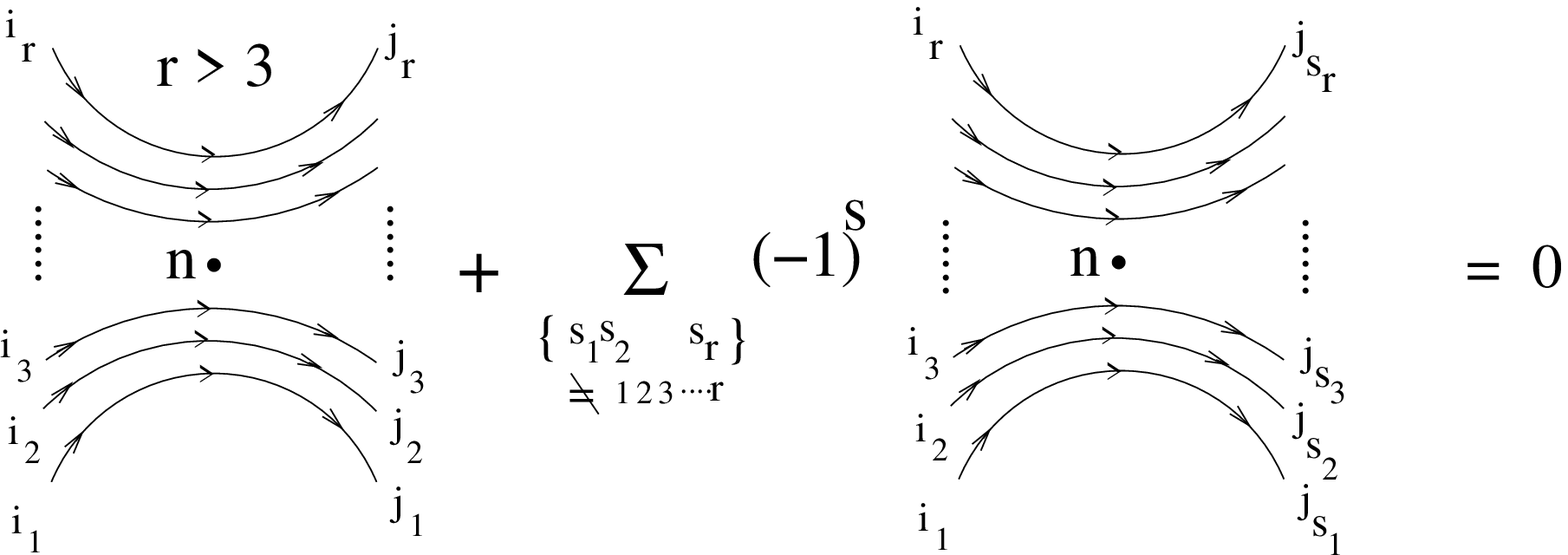}
\end{center}
\caption{The graphical representation of local SU(3) Mandelstam constraints (\ref{nltl}) involving 
only  $L$ type of vertices at lattice site n.} 
\label{su3mcp2}
\end{figure}

\noindent As in the loop formulation where various loop states are mutually related by Mandelstam constraints, 
not all states 
in (\ref{su3gis}) are linearly independent. Infact,  in the present SU(3) prepotential formulation (like in 
SU(2) case) the  Mandelstam constraints become local and take very simple forms in terms of the SU(3) gauge 
invariant vertices in (\ref{gi1},\ref{gi2}) and (\ref{gi3}) at every lattice site n. We start with the simplest 
SU(3) Mandelstam constraints:  
\bea 
A_{[i_1,i_2,i_3]}B_{[j_1,j_2,j_3]} \equiv 
\sum_{\{s_1,s_2,s_3\} \in S_3} (-1)^s  L_{[i_1j_{s_1}]} L_{[i_2j_{s_2}]} L_{[i_3j_{s_3}]}. 
\label{su3mc1} 
\eea
\begin{figure}[h]
\begin{center}
\includegraphics[width=0.8\textwidth,height=0.2\textwidth]
{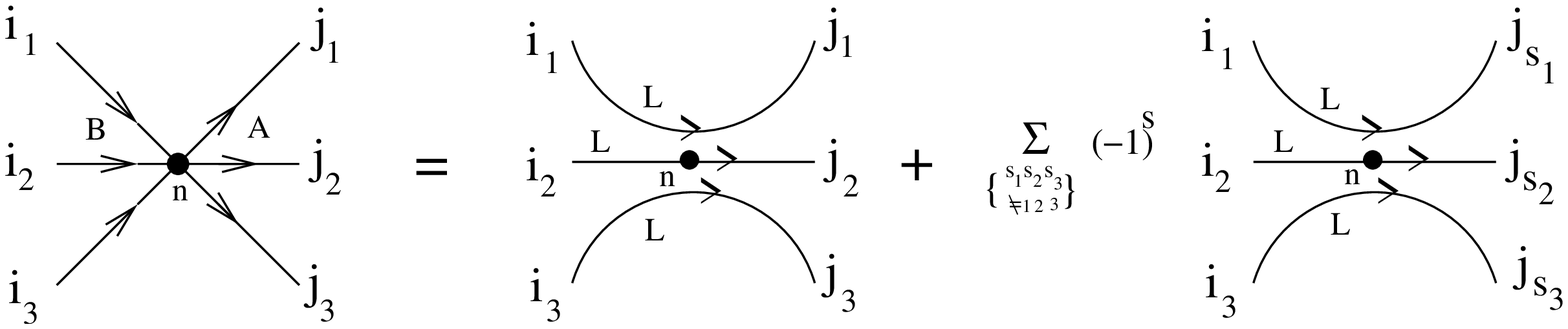}
\end{center}
\caption{The graphical representation of local SU(3) Mandelstam constraints (\ref{su3mc1}) in terms of SU(3) 
gauge invariant vertices $A,B$ and $L$  constructed out of the SU(3) irreducible prepotential operators
at a lattice site n.  The A and B type of vertices at n  annihilate each other to produce 
L type of vertices.} 
\label{su3mcp1}
\end{figure}
In (\ref{su3mc1}), $S_3$ denotes the permutation group of order 3, $\{s_1,s_2,s_3\}$ denote the $3!$ 
permutations of $\{1,2,3\}$ and $s$ is the parity of permutation. In other words, 
the Mandelstam constraints (\ref{su3mc1}) state that the  A and B type vertices annihilate each other in 
pairs to produce L type vertices. The constraints (\ref{su3mc1}) are  illustrated in Figure \ref{su3mcp1}.  
Therefore, the SU(3) gauge invariant states of $(L-A-B)$ type in (\ref{su3gis}) 
can always be written either as $(L-A)$ type or as $(L-B)$ type at each lattice site. It is interesting to 
analyze the Mandelstam constraints discussed in \cite{migdal} in terms of SU(3) prepotential operators.  
Following \cite{migdal}, we consider the set of r $( r >  3)$  loops $C_1(n),~C_2(n), \cdots ,C_r(n)$ 
all based at lattice site n. These loops start from n in the direction $i_1,i_2, \cdots i_r$ and 
come back to  n from directions $j_1,j_2, \cdots j_r$ respectively. 
Then the products of these Wilson loops satisfies:  
\bea 
\sum_{{}^{\alpha_{i_1} \cdots \alpha_{i_r}}_{\beta_{j_1} \cdots \beta_{j_r}}}
\epsilon_{\alpha_{i_1}\alpha_{i_2} \cdots \alpha_{i_r}} \epsilon^{\beta_{j_1}\beta_{j_2} \cdots \beta_{j_r}} 
 \left(W(C_1(n)\right)^{\alpha_{j_1}}{}_{\beta_{i_1}} \left(W(C_2(n)\right)^{\alpha_{j_2}}{}_{\beta_{i_2}} \cdots  
 \left(W(C_r(n)\right)^{\alpha_{j_r}}{}_{\beta_{i_r}}  \equiv  0. 
\eea 
Using the identities 
$$\epsilon_{\alpha_{i_1}\alpha_{i_2} \cdots \alpha_{i_r}} \epsilon^{\beta_{j_1}\beta_{j_2} \cdots \beta_{j_r}} 
= \delta^{\beta_{j_1}}_{\alpha_{i_1}}\delta^{\beta_{j_2}}_{\alpha_{i_2}} \cdots \delta^{\beta_{j_r}}_{\alpha_{i_r}}
- \delta^{\beta_{j_1}}_{\alpha_{i_2}}\delta^{\beta_{j_2}}_{\alpha_{i_i}} \cdots \delta^{\beta_{j_r}}_{\alpha_{i_r}}
+ \cdots $$
(\ref{nlmc}) can be written in terms of traces of Wilson loops \cite{migdal}: 
\bea 
Tr W(C_1) Tr W(C_2) \cdots Tr W(C_r) - Tr W(C_1C_2) Tr W(C_3) Tr W(C_4)  \cdots Tr W(C_r)  + \cdots 
=0. 
\label{nlmc}
\eea 
The Mandelstam constraints (\ref{nlmc}) in terms of the link operators represent 
highly non-local constraints as one can always choose  the loops $C_1,C_2, \cdots C_r$ to be 
as large as one wishes. {\it However, in terms of the prepotentials the constraints (\ref{nlmc}) 
become local.} All one has to do is to replace the Wilson loops in (\ref{nlmc}) by the prepotentials 
which are attached to their starting and end points, i.e,: 
\bea 
W(C_s)^{\alpha_{i_s}}{}_{\beta_{i_s}}  \rightarrow L^{\alpha_{j_s}}{}_{\beta_{i_s}}
\equiv B^{\dagger \alpha_{j_s}} A^\dagger_{\beta_{i_s}}, ~~ s=1,2, \cdots ,r.   
\eea  
Note that unlike  non-local Wilson loop $W(C_s)$, the operators $B^{\alpha_{j_s}}$ and $A^\dagger_{\beta_{i_s}}$ 
and hence $L^{\alpha_{j_s}}{}_{\beta_{i_s}}$ are completely defined at lattice site n. Noting that 
$Tr \left( L^{\alpha_{j_s}}{}_{\beta_{i_s}}\right) = L[i_sj_s]$, the non-local Mandelstam 
constraints (\ref{nlmc}) acquire the following simple local form:    
\bea 
\sum_{\{s_1,s_2, \cdots s_r\} \in S_r} (-1)^s  L_{[i_1j_{s_1}]} L_{[i_2j_{s_2}]} 
\cdots L_{[i_3j_{s_r}]} =0 
\label{nltl} 
\eea 
and are illustrated in Figure \ref{su3mcp2}. {\it Note that all the unnecessary details like shapes, sizes and 
lengths of the loops $C_1,C_2, \cdots C_r$ in (\ref{nlmc}) disappear in  the corresponding prepotential 
form (\ref{nltl})}. 

\subsubsection{\bf The solutions} 

The Mandelstam constraints in their present local prepotential forms (\ref{su3mc1}) and (\ref{nltl}), 
instead of non-local form (\ref{nlmc}) in terms of link operators, are now accessible to explicit local solutions 
like in SU(2) lattice gauge theory \cite{mm}. Note that they  are still infinite in number at 
every lattice site. 
The solutions must be all possible mutually independent linear combinations of the states in (\ref{su3gis}) at a given 
lattice site.  Following the techniques discussed in \cite{sharat1} in the context of duality 
transformations in lattice gauge theories, these linear combinations can be obtained by characterizing 
the resultant states at a site n by their complete SU(3) quantum numbers with the net SU(3) 
fluxes being zero.  This will be SU(3) analogue of SU(2) result (\ref{std2}).  The quantum 
numbers needed to specify such states can be easily computed \cite{sharat1} as follows. 
In d dimension, there 
are 2d links emanating from a lattice site n. 
Each of these 2d directions is attached with SU(3) operators $(A^{\dagger}[i], B^{\dagger}[i], i =1,2, \cdots ,2d)$ 
as shown in Figure \ref{su3s}. Therefore, there are 2d Hilbert spaces associated with a lattice site and  each  
can be characterized by it's SU(3) quantum numbers.  In the standard language \cite{georgi}, the SU(3) irreducible 
representations are completely specified by 5 quantum numbers: $|p,q,i^2,i_z,y\rangle$ where p and q 
are the eigenvalues of two SU(3) Casimir operators and $i, i_z, y$ are the SU(3) ``magnetic" 
quantum numbers representing SU(2) spin, it's third component and hyper charge respectively. 
In the present language with constraints, each of the 2d directions (see Figure \ref{su3s}) 
is associated with 6 harmonic oscillators $(A^{\dagger}[i], B^{\dagger}[i], i =1,2, \cdots ,2d)$ 
and therefore requires 6 occupation numbers to completely specify the basis. The constraints 
$k_{{}_-}[i] \equiv a[i] \cdot b[i] \simeq 0$ reduces this to 5 in each direction.  
Therefore $5 \times 2d = 10d$  
quantum numbers are needed to specify a local Hilbert space basis completely at each lattice 
site. Not all these quantum numbers are independent as 2d of these are related to the previous 
sites by $U(1) \otimes U(1)$ Gauss law constraints (\ref{su3u1c}). Therefore, we are left with $8d$ quantum 
numbers at every lattice site. Finally, the SU(3) gauge invariance further implies 8 constraints. 
Therefore, the net independent quantum numbers are $8(d-1)$ per lattice site. As expected, this is the 
the number of transverse degree of freedom of 8 SU(3) gluons in d dimension at every lattice site. 
The abelian $U(1) \times U(1)$ 
fluxes over the links will now glue these local SU(3) invariant orthogonal basis at neighboring lattice sites 
according to their Gauss laws (\ref{su3u1c}). This will give complete solutions of all the SU(3) Mandelstam 
constraints like what  what was done in SU(2) lattice gauge theory \cite{mm}. 
Infact, the addition of fluxes in SU(3) lattice gauge theory has been discussed in \cite{agms}). 
These results combined with the results of this work should enable us to solve SU(3) Mandelstam 
constraints completely in terms  of vertex operators of section (4.6). 
This explicit construction of all the independent SU(3) loop states 
and their dynamics along the line of \cite{mm}  is in progress and will be reported elsewhere.  

\section{Summary and discussion}

In this work we analyze SU(3) lattice gauge theory in terms of the prepotential 
operators which under gauge transformations transform like fundamental matter fields. 
We constructed the SU(3) irreducible prepotential operators which acting on strong coupling 
vacuum directly created the QCD fluxes around lattice sites.  All SU(3) gauge invariant 
vertices in terms of these QCD flux operators were constructed at every lattice site. 
These SU(3) invariant vertices, in turn, enabled us to cast all SU(3) Mandelstam constraints 
in their local forms. As mentioned in the text this is an essential step towards 
their complete solution.
The complete solution of Mandelstam constraints, 
in turn, will allow us to write down SU(3) lattice gauge theory completely and exactly 
in terms of minimum essential gauge invariant loop and string co-ordinates without any 
redundant loop/strings degrees of freedom. The prepotential operators also allow us to 
simplify lattice gauge theory Hamiltonian as given in (\ref{ham}). In particular, for the present 
SU(3) case, one can simply replace the plaquette or magnetic term  $ Tr U_{plaquette}$  
in (\ref{ham}) by a new plaquette interaction consisting of the 4 L type vertices at 
the 4 corners of every plaquette. Note that the new Hamiltonian constructed this was 
has exactly the same symmetries as (\ref{ham}) and therefore expected to be in the 
same universality class.  The addition of matter field interactions in the prepotential 
formulation is trivial as matter and prepotential have similar SU(3) gauge transformation 
properties. The difference lies in the abelian $U(1) \otimes U(1)$ transformations under 
which matter fields remain invariant.  

The results in this work can also be generalized to SU(N) lattice gauge theory. We can use 
SU(N) Schwinger bosons \cite{md2} or prepotentials to construct SU(N) electric fields on lattice
similar to (\ref{sb}) and (\ref{su3ef}). We need to elevate these prepotentials so that they 
have symmetries of SU(N) Young tableaues inbuilt. As in section (4.6), the SU(N) Mandelstam constraints 
will again be local and can be solved using the techniques discussed in this work. The work in this direction 
is in progress and will be reported elsewhere.   

\noindent{\bf Acknowledgment}  One of the authors (M.M.) would like to thank H. S. Sharatchandra 
for many interesting discussions during the course of this work. 

\appendix 

\section{The projection operators in ${\cal H}_p$:} 

 \noindent In this appendix we briefly discuss the construction of projection operators 
which project ${\cal H}_p$ to ${\cal H}_g$ on every link: 
\bea 
{\cal P} ~\left\{{\cal H}_p\right\}_{link}    =  \left\{{\cal H}_p(\rho=0)\right\}_{link} = 
\left\{{\cal H}_g\right\}_{link}.     
\label{ptg} 
\eea 
The group theoretical details of this construction can be found 
in \cite{ima}.  It is convenient to first break up $\left\{{\cal H}_p\right\}_{link}$ into  
Hilbert spaces containing p (q) quarks and q (p) anti-quark prepotentials on the left (right): 
\bea 
\left\{{\cal H}_p\right\}_{link} = \sum_{p,q=0}^{\infty} \oplus \left\{{\cal H}_p\right\}_{link}(p,q).  
\label{mhs} 
\eea 
These subspaces $\left\{{\cal H}_p\right\}_{link}(p,q)$ are themselves direct products of 
left and right Hilbert spaces: 
\bea 
\left\{{\cal H}_p\right\}_{link}(p,q) = \left\{{\cal H}_p^L\right\}_{link}(p,q) \otimes
\left\{{\cal H}_p^R\right\}_{link}(q,p). 
\eea 
The basis vectors spanning $\left\{{\cal H}_p^l\right\}_{link}(p,q), l= L,R$ are given in 
terms of left and right flux creation operators: 
$$\hat{L}_{\alpha_1\alpha_2 \cdots \alpha_p}^{\beta_1 \beta_2 \cdots \beta_q} |0\rangle_L 
~ \equiv ~  a^{\dagger}_{\alpha_1}(L) \cdots  a^{\dagger}_{\alpha_p}(L) b^{\dagger\beta_1}(L) \cdots  
b^{\dagger\beta_q}(L)|0\rangle_L,$$   
and 
$$\hat{R}_{\gamma_1\gamma_2 \cdots \gamma_q}^{\delta_1 \delta_2 \cdots \delta_p} |0\rangle_R 
~ \equiv ~ a^{\dagger}_{\gamma_1}(R)  \cdots  a^{\dagger}_{\gamma_q}(R)  b^{\dagger\delta_1}(R) \cdots   
b^{\dagger\delta_p}(R) |0\rangle_R$$   
in (\ref{su3lst}). 
We now construct the projection operators ${\cal P}_l(p,q)$ in each of these subspaces with 
\bea 
{\cal P} = \sum_{p,q=0}^{\infty} \oplus {\cal P}_L(p,q) \otimes {\cal P}_R(q,p). 
\label{pol} 
\eea 
The  left and right projection operators ${\cal P}_l(p,q), l=L,R$ are of the 
form \cite{ima}: 
\bea
{\cal P}_L(p,q) & \equiv & \sum_{r=0}^{\infty} g_r(p,q) \left(k_+(L)\right)^r \left(k_-(L)\right)^r 
\nonumber \\
{\cal P}_R(q,p) & \equiv & \sum_{r=0}^{\infty} h_r(q,p) \left(k_+(R)\right)^r \left(k_-(R)\right)^r 
\label{abc} 
\eea
The unknown coefficients g and h in (\ref{abc}) are fixed by demanding the Sp(2,R) constraints 
(\ref{kan}): 
\bea 
k_-(L) {\cal P}_L(p,q) \left\{{\cal H}_p^L\right\}_{link}(p,q) = 0, ~~~~ 
k_-(R) {\cal P}_R(q,p) \left\{{\cal H}_p^R\right\}_{link}(q,p) = 0.  
\label{aab} 
\eea 
The solutions of the equations (\ref{aab}) are \cite{ima}: 
\bea
g_r(p,q)= h_r(q,p) = 
\frac{(-1)^r}{r!}\frac{( p+q+1-r)!}{( p+ q+1)!},
\label{xbv}
\eea
leading to:
\bea
{\cal P}_L{(p,q)} & = & \frac{1}{(p+q+1)!} \sum_{r=0}^{\infty}
\frac{(-1)^r}{r!}({p}+{q}+1-r)! \left(k_+(L)\right)^r \left(k_-(L)\right)^r, \nonumber \\
{\cal P}_R{(q,p)} & = & \frac{1}{(p+q+1)!} \sum_{r=0}^{\infty}
\frac{(-1)^r}{r!}({p}+{q}+1-r)! \left(k_+(R)\right)^r \left(k_-(R)\right)^r. 
\label{P0}
\eea
Note that the SU(3) irreducible prepotentials in (\ref{modsb2}) already commute with 
the Sp(2,R) constraints (\ref{all0}) and therefore acting on the strong coupling vacuum 
directly generate the gauge theory Hilbert space ${\cal H}_g$. In other words: 
\bea 
\hat{{\cal L}}_{~\alpha_1\alpha_2 \cdots \alpha_p}^{~\beta_1 \beta_2 \cdots \beta_q}|0\rangle_L  
 = {\cal P}_L(p,q) \hat{{L}}_{~\alpha_1\alpha_2 \cdots \alpha_p}^{~\beta_1 \beta_2 \cdots \beta_q}
|0\rangle_L, ~~~~  
\hat{{\cal R}}_{~\gamma_1\gamma_2 \cdots \gamma_q}^{~\delta_1 \delta_2 \cdots \delta_p}|0\rangle_R.  
= {\cal P}_R(q,p) \hat{{R}}_{~\gamma_1\gamma_2 \cdots \gamma_q}^{~\delta_1 \delta_2 \cdots \delta_p}|0\rangle_R  
\label{cptp} 
\eea  
are the relations amongst the SU(3) reducible and irreducible flux operators on the left and right 
side of every link. 

\section{The electric field constraints} 
 
Using the $\lambda$ matrix identity: 
$\sum_{a=1}^{8} \left(\frac{\lambda^a}{2}\right)^{\alpha}_{\beta}  
\left(\frac{\lambda^a}{2}\right)^{\gamma}_{\sigma} = \frac{1}{2} 
\delta^{\alpha}_{\sigma} \delta^{\gamma}_{\beta} - 
\frac{1}{6}  \delta^{\alpha}_{\beta} \delta^{\gamma}_{\sigma}$, the squares of 
left and right electric fields can be written as:    
\bea 
\sum_{a=1}^{8} E_L^a(n,i)  E_L^a(n,i) 
& = & \hat{N}(L)\left( \frac{\hat{N}(L)}{3} +1\right)   + 
\hat{M}(L)\left( \frac{\hat{M}(L)}{3} +1\right) -~ 
k_+(L)k_-(L) + \frac{1}{3} \hat{N}(L) \hat{M}(L) \nonumber \\  
\sum_{a=1}^{8} E_R^a(n,i)  E_R^a(n,i) 
& = & \hat{N}(R)\left( \frac{\hat{N}(R)}{3} +1\right) 
+ \hat{M}(R)\left( \frac{\hat{M}(R)}{3} +1\right) - 
k_+(R)k_-(R) + \frac{1}{3} \hat{N}(R) \hat{M}(R). \nonumber   
\label{lre} 
\eea   
The electric field constraints (\ref{E2=e2}) along with the 
$U(1) \otimes U(1)$ Gauss law constraints (\ref{su3u1c}) imply: 
\bea 
k_+(L)k_-(L) = k_+(R)k_-(R). 
\label{aee1} 
\eea 
On the other hand, the action of $k_+k_-$ on a general 
Sp(2,R) irrep. $|k, m \rangle$ is given by \cite{nm}: 
\bea 
k_+k_-|k, m \rangle = (m-k) (m+k-1) |k, m \rangle,   
\label{aee2} 
\eea 
where, $m = k + \rho $. In the present case the electric field constraint (\ref{lre}) and the eigenvalue 
equation (\ref{aee2}) imply: 
\bea 
\left(m(L) - k(L)\right) \left(m(L) + k(L) -1\right)  = \left(m(R) - k(R)\right) \left(m(R) + k(R) -1\right). 
\label{aee3} 
\eea 
As  $k(L) = k(R) = \frac{1}{2}(p+q+3)$, we get the unique solution of (\ref{aee3}): 
\bea 
\rho_L(n,i)  =\rho_R(n+i,i). 
\eea 
Therefore, in the prepotential Hilbert space ${\cal H}_p$ the left and the right Sp(2,R) 
``magnetic" quantum numbers are same on every link.

\end{document}